\pdfoutput=1   
\documentclass[]{aa}

\usepackage[english]{babel}
\usepackage{csquotes}
\usepackage{url}
\usepackage{enumitem}
\usepackage{bm}
\usepackage{xcolor}
\usepackage{afterpage}
\usepackage[normalem]{ulem}
\usepackage{graphicx}
\usepackage{nicefrac}
\usepackage{stfloats}
\usepackage{txfonts}
\usepackage{amssymb}
\usepackage{amsmath}
\usepackage{siunitx}
\usepackage{booktabs}
\usepackage[font=small]{caption}
\usepackage{balance}
\usepackage{ragged2e}
\usepackage[margincaption]{sidecap}
\usepackage{hyperref}

\hypersetup{colorlinks = true,
            linkcolor = blue,
            urlcolor  = blue,
            citecolor = blue,
            anchorcolor = blue}

\newcommand{\teff}{T$_\mathrm{eff}$}
\newcommand{\Rsun}{R$_\odot $}
\newcommand{\Msun}{M$_\odot $}

\newcommand{\Mj}{M$_{J}$}

\newcommand{\secname}{Sec.}
\newcommand{\eqname}{Eq.}
\makeatletter
\renewcommand*\aa@pageof{, page \thepage{} of \pageref*{LastPage}}
\makeatother

\begin{document}

    \title{Statistics of Magrathea exoplanets beyond the Main Sequence}
    \subtitle{Simulating the long-term evolution of circumbinary giant planets with \texttt{TRES} }

    \author{
        G. Columba\inst{\ref{unipd},\ref{oapa}}
        \and
        C. Danielski\inst{\ref{iaa}}
        \and
        A. Dorozsmai \inst{\ref{unibirm}}
        \and
        S. Toonen\inst{\ref{api}}
        \and
        M. Lopez Puertas \inst{\ref{iaa}}
        }

    \institute{
        Department of Physics and Astronomy "Galileo Galilei" (DFA), University of Padua, via Marzolo 8, 35131 Padua, Italy.\label{unipd} \\
        \email{gabriele.columba@phd.unipd.it}
        \and
        INAF – Osservatorio Astronomico di Padova, Vicolo dell’Osservatorio 5, 35142 Padua, Italy.\label{oapa} 
        \and
        Instituto de Astrof\'isica de Andaluc\'ia, CSIC, Glorieta de la Astronom\'ia s/n, 18008 Granada, Spain.\label{iaa}
        \and
        Institute of Gravitational Wave Astronomy and School of Physics and Astronomy, University of Birmingham,\\ Edgbaston, Birmingham B15 2TT, United Kingdom \label{unibirm}
        \and    
        The Anton Pannekoek Institute for Astronomy, University of Amsterdam (UvA), Science Park 904, 1098 XH Amsterdam, Netherlands.\label{api} 
    }

   \date{}

 
  \abstract
    {Notwithstanding the tremendous growth of the exoplanetary field in the last decade, limited attention has been paid to the planets around binary stars, which represent a little fraction of the total discoveries to date. 
    Circumbinary planets (CBPs) have been discovered primarily with transit and eclipse timing variation methods, mainly around Main Sequence (MS) stars. No exoplanet has been found orbiting double white dwarf (DWD) binaries yet.}
    {
    In the interest of expanding our understanding on the final fate of CBPs, we modelled the long-term evolution of these circumbinary systems, throughout the life stages of their hosts, from MS to WD. Our goal is to provide the community with both theoretical constraints on the evolution of CBPs beyond the MS, and the occurrence rates of planet survivals, ejections and collisions throughout the ageing of the systems. 
    }
    {We further developed the publicly available Triple Evolution Simulation (\texttt{TRES}) code, to adapt it to the mass range of sub-stellar objects (SSO). We did so by implementing a variety of physical processes that affect giant planets and brown dwarfs. 
    We used \texttt{TRES} to simulate the evolution, up to one Hubble time, of two synthetic populations of circumbinary giant planets. Each population has been generated using different priors for the planetary orbital parameters. 
    }
    { In our simulated populations we identified several evolutionary categories, such as survived, merged, and destabilised systems. Our primary interest is those systems in which the planet survived the WD formation of the stars in the binary. We name such planets \emph{Magrathea}. 
    We found that a significant fraction of simulated CBPs survive the entire system evolution and orbit a DWD binary. 
    The planetary mass does not play a role in their survival. In the absence of effective inward migration mechanisms, this category of planets is characterised by long periods.
    }
    {\emph{Magrathea} planets are a natural outcome of triple systems evolution and our study indicates that they should be relatively common in the Galaxy. These gas giants can survive the death of their binary hosts if they orbit far enough to avoid engulfment and instabilities. 
    Our results can ultimately be a reference to orient future observations of this uncharted class of planets and to compare different theoretical models. 
    }


   \maketitle
%

\section{Introduction}
    \label{sec:intro}

    In spite of the diversity of exoplanets uncovered in the last few years, the known sample of planetary bodies does not uniformly cover the stellar H-R diagram. 
    Several H-R regions, that show evidence of the possible existence of planets, 
    show little statistics mostly due to observational bias.
    For instance, while the majority of the planets has been found around Main Sequence (MS) stars, only one confirmed planet \citep{Blackman21} and several candidates \citep{Gansicke2019,Vanderburg20,Arenou2022:gaia} have been identified orbiting single white dwarfs (WDs). These detections arrived after many years of dedicated research (see \citealt{Veras2021} and references there-in for a recent review). 
    Another offbeat planetary population is the circumbinary one, the so-called \textquote{P-type} systems\footnote{ We refer to \cite{MarzariThebault} for a thorough overview on the characteristics of both P-type and  S-type populations}, where the planet orbits a compact binary star with typical inner orbital separations of less than $\sim$ 10\,au.
    With the years, multiple studies showed that stars in binary and multiple systems are ubiquitous throughout the Milky Way \citep{Raghavan2010,DucheneKraus2013,Tokovinin2014, MoeDistefano2017,Tokovinin21}. Such a physical feature naturally sparkled the interest of planets-chasers, who sky-hunted planets orbiting multiple star systems through various methods (i.e., astrometry 
    e.g., \citealt{Sahlmann2015}, 
    transit e.g., \citealt{Doyle2011,Kostov2014,Kostov2020,MartinFabrycky2021}, 
    and radial velocities e.g., \citealt{Martin2019:bebop,Triaud2022,Standing2023}), and who also explained the small statistics due to selection bias. 
    For instance, when focusing on the P-type population around double MS binaries, most of the evidence has been collected through the transit method, which highlighted the coplanarity of planetary and binary orbits. However, \cite{Armstrong2014} showed that if planetary orbital inclinations were randomly distributed with respect to the binary orbital plane, then the inferred frequency of planets in circumbinary orbits should be exceptionally high compared to that around single stars. Yet, if the distribution favoured coplanarity, then the frequency of CBPs would be consistent with what has been observed for single stars' planets. But if CBPs were to have at least the same occurrence rate of most of the 5300 planets discovered until now, why cannot we see them? The answer most probably lies in the parameter space that this population covers, which makes their detection challenging with the instruments and observational sampling we currently use.
    
    Shifting now the attention to the H-R diagram region of evolved stars, very recent works showed that the stellar multiplicity percentage is held there too. For instance, thanks to the exceptional Gaia DR3 data \citep{GaiaDR3}, a new sample of white dwarfs in binary systems have been established \citep{Arenou2022:gaia}. Similarly, a more focused work on the hot WD sample identified a probable binary fraction of $\sim 55\%$ \citep{GomezMunoz2022}. For both studies the nature of the stellar companion is though uncertain.\\ 
    However, very little is known about exoplanets about evolved binary systems.
    So far, among the S-type systems (\citealt{MarzariThebault} and references there-in) planets have only been found orbiting the MS component, never the WD component, (e.g., 
    HD 8535 - \citealt{Naef2010,Mugrauer2019}; 
    WASP-98 - \cite{Hellier2014,Southworth2020}; 
    HIP 116454 - \citealt{Vanderburg2015ApJ:HIP116454}; 
    TOI-1259 - \citealt{Martin2021};
    TOI-3714 - \citealt{Canas2022}). 
    In these type of systems the stellar binary is usually very wide (e.g., 302\,au for TOI-3714 and 3500\,au for WASP-98), and therefore the evolution of the stars is not affected by binary interactions.
    This means that the planet, unless it hopped from the evolved primary star \citep{KratterPerets2012}, has not directly suffered the evolution of its host star, which is still in the MS stage.
    
    On the other hand, several P-type exoplanets are known to orbit a binary where one of the stars evolved off the MS.
    Among the confirmed $\sim 45 $ circumbinary SSOs known to-date, 7 planets orbit one WD star (for a total of 5 planetary systems: DP Leo \citep{Beuermann+2011:DPleo}, NN Ser \citep{Beuermann+2010:NNser}, PSR B1620-26 \citep{Sigursson2003:PSRb1620}, RR Cae \citep{Qian+2012:RRcae}, UZ For \citep{Potter+2011:UZfor}); and 8 planets have one host that has completed a giant branch phase, but it is not yet a WD (for a total of 5 systems: Kepler-451 \citep{Esmer+2022:Kep451}, HW Vir \citep{Beuermann+2012:HWvir}, MXB 1658-298 \citep{Jain+2017:MXB1658}, NSVS 14256825 \citep{Zhu+2019:NSVS1425}, NY Vir \citep{Song+2019:NYvir})\footnote{The list of confirmed circumbinary planets has been obtained from the \href{https://exoplanetarchive.ipac.caltech.edu}{NASA Exoplanets Archive}, 
    with integrations and cross-checks from 
    \href{http://exoplanet.eu}{The Extrasolar Planets Encyclopaedia} and 
    the \href{https://github.com/OpenExoplanetCatalogue/open_exoplanet_catalogue}{Open Exoplanets Catalogue}}.
    Detecting planets around evolved systems is usually through an indirect method: the direct observation of WDs by optical inspection is known to be quite hard due to their small radii and faint total luminosity, which inevitably reflects on the hurdle of detecting a possible planetary companion with a high significance.
    The typical technique used to discover the known evolved P-type systems is the eclipse transit timing variation (ETTV) method, where the significance of a detection increases with the length on the observational baseline. In fact, in some cases the relatively short baseline has been source of discussions on the presence and/or nature of some of these exoplanets (e.g., \citealt{Bours2016}). This is mostly due to confusion with the Applegate mechanism \citep{Applegate1992}, or questions on the system stability (e.g., \citealt{Wittenmyer2013}), or the necessity to refine the ETTV models \citep{Pulley2022}, as alternative interpretations of the data.\\
    When comparing these systems to the S-type siblings, the stellar binaries presents compact orbits,  meaning that any stellar evolution is affected by binary interactions (such as mass transfer episodes), meaning that their planetary companions must have experienced, and survived, the consequences of at least one binary evolutionary step.
    
    The evidence of a larger number of planetary systems in evolved binaries than single-WDs, has been argued not to be a coincidence. A work by \cite{Kostov16:CE} studied the dynamical evolution of nine \textit{Kepler} CBPs, with particular focus on the common envelope (CE) phases of the stars. The authors found that the CBPs have more chances of survival when orbiting compact binary systems than single stars of similar mass. As an example, a planet orbiting a binary of total mass $\simeq 1.3 $ \Msun\ at the distance of Mercury from the Sun, would likely survive the giant phases of its hosts, whereas Mercury, and Earth too, are going to be completely engulfed as soon as the Sun  radius expands along the red giant branch \citep{SchroderSmith2008}.
    
    The next evolutionary step of these systems is the evolution of the secondary stellar companion, which by definition has a lower mass than the mass of the primary progenitor. By simplifying (for benefit of the reader), this process can either results in the merger of the two stars, or in a double degenerate stellar system whose components could be either a double white dwarf (DWD) or a WD and a neutron star. To date about a hundred  double generate systems have been detected, using a variety of methods \citep[see][for a recent overview]{Korol2022}. The majority of sources have been detected by the Supernova Ia Progenitor surveY (SPY) \citep[][and references therein]{Napiwotzki2020} and the Extremely Low-Mass (ELM) survey \citep{Brown2010, Brown2020} which is based on the Sloan Digital Sky Survey. In addition, the ZTF high-cadence Galactic plane survey \citep{Masci19} has been particularly effective in detecting extremely compact DWD systems \citep{Burdge2019, Burdge2020}. 
    All these surveys have helped increasing the number of known short period binaries, yet the largest improvement in the field will come with the Laser Interferometer Space Antenna (LISA, \citealt{Amaro-Seoane17:LISA}), which is expected to find around $\sim 10^4$ DWDs \citep{Korol2017,Lamberts2019, Breivik2020,Korol2022} and many other types of degenerate short period binaries \citep{LISAwhitepaper}.

    The LISA mission will be transformative not only for the stellar physics of binaries but also for exoplanets, being the only planned mission with the potential to detect giant planets around DWDs anywhere in the Milky Way \citep{TamaniniDanielski19:CBP,Danielski19:CBP} and in the Large Magellanic Cloud \citep{Danielski20:CBP}.
    At present no exoplanets orbiting DWDs have been found, but the results of \cite{Kostov16:CE}, coupled with the fact that 97$\%$ of existing stars will eventually end their life as WDs \citep{Fontaine2001}, suggest that DWD binaries will likely harbour surviving planets. Our work aims to thoroughly explore such a possibility via modelling of the long-term evolution of different populations of CBPs.
    
    The focus of this work is to quantify the survival rate of circumbinary planets in the context of the host-binary evolution, from the Zero Age Main Sequence (ZAMS) to one Hubble time i.e., the age of the Universe. 
    For our purposes, we exploited the publicly available Triple Evolution Simulation code (\texttt{TRES}\footnote{ \url{https://github.com/amusecode/TRES})}, \citealt{Toonen16:tres}, see also \secname~\ref{sec:tres}) and expanded it into the \texttt{TRES-Exo} code, 
    a \texttt{TRES} option dedicated to the evolution of compact binaries hosting a single giant planet. 
    We then used \texttt{TRES-Exo} to produce two populations of systems with different initial planetary priors, to evaluate their impact on the final populations and on the occurrence rates. 
    An important collateral science to our analysis is the exploration of the possibility of using the surviving planets (which we refer to as {\it Magrathea}\footnote{From ``The Hitchhiker’s Guide to the Galaxy'' book \citep{Magrathea} 
    Magrathea is a planet orbiting a binary star burning with “white fire”, which we conveniently interpreted to be two white dwarfs, since they emit white light instead of yellow, like our Sun in the visible band.
    }
    planets) for constraining the CE phase of the binary \citep{Ivanova2013}.
    
    Our work is set within the broader context of study for the development of the planetary detection science case of the LISA mission. Such a development includes the LISA detection prospects of SSOs orbiting DWDs through Bayesian analysis by \cite{Katz2022}, the determination of the LISA detection efficiency by \cite{Danielski-in-prep}, and the study of second-generation planetary formation around DWD by \cite{Ledda2023}.
    
    This paper has the following structure: in \secname~\ref{sec:models} we present the physical models employed to simulate the circumbinary SSOs and implemented in \texttt{TRES}.
    In \secname~\ref{sec:popsyn} we illustrate the initial properties of our synthetic populations. In \secname~\ref{sec:results} we present our results, divided by category of evolution, and we highlight the most important final properties of the synthetic sample. In \secname~\ref{sec:discuss} we discuss some implications of our results and we draw our conclusions in \secname~\ref{sec:conclude}.

\section{Methods}
    \label{sec:models}

    \subsection{The Triple Evolution Simulation code}
        \label{sec:tres}
    
        The powerhouse of our simulations is the Triple Evolution Simulation package (\texttt{TRES}), which is publicly available on github and described in depth in \cite{Toonen16:tres}.
        This software was designed to simulate hierarchical triple star systems, and it includes a thorough analytical treatment including secular orbital evolution, and various stellar interactions (e.g. tides, common envelope evolution, stellar winds, supernovae and associated natal kicks).
        The single stellar evolution is modelled by \texttt{SeBa}  \citep{Nelemans01:DWD,Toonen12:DWD,Toonen&Nelemans13:CE}, which is a parameterised, fast stellar evolution code. \texttt{SeBa}  determines the most important stellar parameters (such as mass, core mass, radius, and luminosity) at a given timestep, using the fitting formula of \citet{Hurley00}. 
        We have further developed \texttt{TRES} and {SeBa} so that we can simulate substellar bodies as well, namely brown dwarfs and gaseous planets.
        
        In the following section we illustrate the key physical models that were deemed meaningful for our simulations.

    \subsection{Developing \texttt{TRES} for planets: \texttt{TRES-Exo}}
        \label{sec:newprocesses}
    
    Given that our main goal is to explore the fate of exoplanets in hierarchical triple systems, we added new features to \texttt{TRES} that reflects the physical processes and properties typical of the giant planets mass range (0.2 \Mj\ -- 16 \Mj). We describe each feature in the following paragraphs. It should also be noted that we leave the modelling of brown dwarfs long-term evolution for future analysis, and we focus here on the planetary mass range.

    \subsubsection{Stability criteria}
        \label{subsec:stability}

        Being interested in the long-term evolution of systems, it is relevant to evaluate the stability of the simulated triples. New stability criteria have been added to \texttt{TRES} to choose between. For our purpose of generating circumbinary planets, in particular, we have decided to adopt the criterion of \cite{Holman&Wiegert99:stability} for P-type stability, which formulates the critical minimum semi-major axes ratio as:
            \begin{equation}\label{eq:stability}
                \resizebox{1\hsize}{!}{$\frac{ a_{\rm out} }{ a_{\rm in} } \Bigr|_{\rm crit} = 1.6 + 5.1 e_{\rm in} - 2.22 e_{\rm in}^2 + 4.12 \mu - 4.27 e_{\rm in} \mu -5.09 \mu^2 + 4.61 e_{\rm in}^2 \mu^2$}
            \end{equation}
        where $e_{\rm in}$ is the eccentricity of the inner binary and $\mu = \frac{m_2}{m_1 + m_2}$ its mass ratio.
        This criterion, used also in recent N-body literature (see, e.g., \citealt{Ballantyne21:stability}), results in a minimum planetary semi-major axis of a$_{\rm out}$ = $ 2.39 \ a_{\rm in} $, for the special case of an equal mass inner binary on circular orbits; higher eccentricity causes the closest stable orbit to move farther away.
        Note that the critical outer semi-major axis depends on the parameters of the inner binary alone.

    \subsubsection{Tidal interactions}
    
        Tidal forces are responsible for several phenomena in binary and close-in systems, including: orbit circularisation, synchronisation of rotational periods of the bodies and shrinking of the orbital distances; specifically for eccentric orbits, they are also known to cause apsidal precession, i.e. a secular change in the position of a body's periastron.
        In general, the presence of tides translates to torques that exchange angular momentum between the star and the orbit, together with a dissipation of the energy into the tide itself.
        
        \texttt{TRES} includes a treatment of tidal interaction through the model of \cite{Hut81:tides}, which assumes small tides in their equilibrium shape, in a constant time lag with respect to the line of the centres. The outcome of such tidal interaction is to change the orbital parameters of the stars, eventually either approaching an equilibrium configuration - a circular binary of stars rotating synchronously with their orbital period - or else leading them to a fatal decay of their orbits until they merge together. Hut's differential equations for the change in orbital parameters due to tides are all proportional to the $k/T$ factor (the ratio of the apsidal motion constant to the tidal damping timescale).
        
        The particular physical properties of a celestial body correspond to different tidal dissipation mechanisms, influencing the timescales of the tidal effects itself. For stellar objects, \texttt{TRES} includes damping mechanisms for convective, radiative or degenerate stars, following \cite{Hurley02}.
        For giant planets and brown dwarfs, we adopted the tidal timescale used by \cite{FabryckyTremaine2007:tides} (see their Eq. A9, in the Appendix), plugging their constant value of the viscous timescale $ t_V = 0.001 \si{yr} $ and using our interpolated value of the apsidal motion constant $ k_2 $ (described in detail in \secname~\ref{subsec:amc}).

    \subsubsection{Stellar and substellar structure constants}
        \label{subsec:amc}

        We improved the accuracy of our simulations, exploiting the latest results of \cite{Claret19:k2}, to provide more reliable values for two internal structure constants: the gyration radius and the \enquote{Apsidal Motion Constant} ($k_2$). We implemented the new tables for both stars and planets, interpolating the parameters as a function of the system age, rather than adopting fixed average values. 
        The gyration radius is relevant to the computation of the moment of inertia, while the apsidal motion constant is related to the strength of tidal perturbations (and thus the effects described in the previous paragraphs).
        We note that there is a historical ambiguity for the name of the apsidal motion constant, due to different derivations by \cite{Sterne1939:k2}, or \cite{Love1911}, after whom it was also named \enquote{2nd Love Number}, with only a difference by a factor two: $ k_{2,\mathrm{AMC}} = \nicefrac{1}{2} \, k_{2,\mathrm{Love}} $. Such a terminology, the ``Love number'', is though usually employed in the planetary science field only, and not in the stellar physics field.
        
        \cite{Claret19:k2} results were obtained using the MESA package (Modules for Experiments in Stellar Astrophysics; \citealp{Paxton+2011:mesa, Paxton+2015:mesa}), version R10398, and they include the stellar internal structure constants ($k_2$, $k_3$ and $k_4$), the radius of gyration and the gravitational potential energy. Based on this work, where those parameters were computed up to the first ascent giant branch, our updated models now cover a larger range of stellar life: from the pre-main-sequence phase, up to the white dwarf stage, as required by our simulations. We report in \tablename~\ref{tab:claretmodels} a summary of these new models.
        The mass range covered in our study goes from 0.95\,\Msun \ to 10\,\Msun, for three different metallicities: [Fe/H] = 0.00,  -0.50 and -1.00.
        For the mixing length parameter a solar-calibrated value of $ \alpha_{MLT} = 1.84 $ was adopted and microscopic diffusion included. Convective core overshooting was taken into account with the diffusive approximation, using the parameter $ f_{ov} $ and its relationship with the stellar mass, as illustrated by \cite{ClaretTorres19}. For the opacities we adopted the elements mixture by \cite{Asplund09}, whereas the helium abundance follows the relation $ Y = 0.249 + 1.67\, Z $, with "Z" being all elements heavier than He.
 
        \begin{table}[tbh]         
            
            \caption{ Values of the main parameters adopted for the stellar models.}
            \label{tab:claretmodels} 
            \centering
            \begin{tabular}{l c }
                \toprule
                Parameter & Value \\
                \midrule
                Stellar mass range    &   \SIrange{0.95}{10}{M_\sun}   \\
                SSO mass range     &    \SIrange{0.2}{16}{M_{J}}    \\
                \text{[Fe/H]}    &   0.0, -0.5, -1.0    \\
                Mixing length   &   1.84   \\
                \bottomrule
                \end{tabular}
        \end{table}

    \subsubsection{The atmospheric evaporation}
        \label{sec:evap}
    
        Evidence of atmospheric evaporation was revealed not long after the first detections of transiting exoplanets. The main probe (among several) for the existence of outflowing gas is the Lyman-$\alpha$ line absorption during planetary transit \citep{Owen19:evap}. Traces of neutral hydrogen outside the planet's Roche lobe show that its atmosphere extends to unbound regions and is thus able to escape. 
        
        Many processes can cause the gas to flow out, but in this work we have only accounted for the \textit{energy-limited} mass-loss, which, in the context of close-in gaseous planets, proves to be the prevailing process \cite{Owen19:evap}. 
        The latter is a thermal evaporation process: the high-energy flux from the star(s) is absorbed by the upper layers of a planetary atmosphere, to the point where their kinetic energy overcomes the planet's gravitational potential and the gas is free to escape. Therefore, it becomes critical to assess the flux of X-ray and EUV radiation received by the planet along its orbit, depending on the stellar life stage. For the modelling of atmospheric evaporation, we have referred to the work of \cite{Flaccomio03a:XUV, Wright11:XUV, ForcadaMicela14:evap, Schreiber19:WD, Owen19:evap}. 
        The mass-loss rate due to the high-energy flux $F_\mathrm{XUV}$, also known as photoevaporation, can be computed as in \cite{Owen19:evap}:
        \begin{equation}
            \dot M = \eta \frac{\pi R_p^3 F_{\mathrm{ XUV}}}{G M_p K_{\mathrm{ Erk}}}
        \end{equation}
        
        \noindent
        where $\eta$ is an energy-transfer efficiency factor and $K_{\mathrm{Erk}}$ is a term accounting for the fact that the gas does not need to reach infinity, to escape the gravitational well of the planet, since it can be considered "lost" as soon as it surpasses its Roche lobe radius \citep{Erkaev07:evap}. Recent studies have found the $\eta$ factor to have a value around 0.2 or lower \citep[e.g.,][]{PenzMicela08:XUV, Lammer+2009:evap, Lampon2023}. For our analysis we adopt $\eta$ = 0.2 as upper limit \citep{Owen19:evap}.
        
        To compute the mean XUV flux a planet receives from the binary, we average on one planetary orbital period, to be consistent with the secular approximation (note that \texttt{TRES} does not keep track of the orbital phases at each time):
        \begin{equation}
            F_{\mathrm{ XUV}} = \frac{1}{4 \pi P_{\mathrm{ orb}}} \int_{0}^{P_\mathrm{ orb}} \left( \frac{L_1}{d^2_1} + \frac{L_2}{d^2_2} \right) \, dt
        \end{equation}
        where $L_1$ and $L_2$ are the XUV luminosities of the two host stars and $d_1, d_2$ their respective three-dimensional distances from the planet, time-dependent variables. 
        Both $d_1, d_2$ are dependent on the stellar phase, the planetary phase and the relative inclination between the two orbital planes. 
        We compute the average distance of a planet from the system's centre of mass as $ \Bar{d} = a ( 1 + e^2 / 2 ) $, which is the time-averaged distance in an orbit with semi-major axis $ a $ and eccentricity $ e $ \citep[e.g.][]{Williams2003:orbits}.
        From trigonometric decomposition, we then compute $d_1, d_2$ as a function of $ \Bar{d} $.
        
        The luminosity of stars in the high-energy band is not a well-characterised quantity to plug into the equations. Until now, in fact, the main focus of most photo-evaporation studies has been the high-energy emission of Main Sequence (MS) or pre-MS stages, but the evidence of planets around evolved stars makes it necessary to extend these studies to all kinds of stars, up to the WD stage. We have used different prescriptions to account for the evaporation of planetary atmospheres caused by stars with a range of different masses and varying life stages. 
        
        We employed the results of \cite{ Wright11:XUV}, concerning the relationship between stellar activity and rotation, to assign the X-ray emission to low-mass stars. The authors have performed a fit to observational data and found two regimes of emission: saturated and unsaturated, based on the Rossby number $ Ro = P_{\mathrm{ rot}} / \tau $, where $P_{\mathrm{ rot}}$ is the rotation period of the star and $ \tau $ its convective turnover time. Their best fit are:
        
        \begin{equation}
            \frac{L_X}{L_{\mathrm{ bol}}} = R_{\mathrm X} =
            \begin{cases}
                R_{\mathrm{ X, sat}}               & Ro \leq Ro_{\mathrm{ sat}} \\
                C \cdot Ro^\beta         & Ro > Ro_{\mathrm{sat}} \\
            \end{cases}
        \end{equation}
        with $ R_{\mathrm{ X, sat}} = 10^{-3.13} $, $ \beta = -2.70 $ and $ Ro_{\mathrm{sat}} = 0.13 $, the threshold of rotational period above which the stars saturate the emission and which is independent of stellar type. The behaviour of stars following this model is generally flat at saturation value at early phases in their life, when they have faster rotational velocities, and then it decays with time proportionally to their rotation. To compute the Rossby number we also took advantage of the analytical fit provided by \cite{Wright11:XUV} for the convective turnover time:
        \begin{equation*}
            \log( \tau) = 1.16 - 1.49 \log(M/ \mathrm{ M_\odot}) - 0.54 \log^2 (M/ \mathrm{ M_\odot}).
        \end{equation*}
        
        For stars with a mass below 2\,\Msun  we used this model to estimate the X-ray emission during their entire life before the WD stage. For stars with $ 2 < M/\mathrm{M_\odot} < 3 $ we used this model only during the post-MS phases, whereas for their MS we imposed the value of $ R_{\mathrm X} \approx 10^{-3.5} $, taken from \cite{Flaccomio03a:XUV}. The extension of the Rossby number approach for the X emission of stars beyond the main sequence is an approximation necessary to cover the lack of information on this topic. Although far from exact, this approach appears to be a plausible solution: \cite{Dixon20} conducted a study on red giants and highlighted the remarkable similarity between these evolved stars and M-dwarfs emissions. In their sample they found, in fact, a correlation of the stellar activity, in terms of NUV excess, to the rotation, suggesting a possibly analogue mechanism for the stellar dynamo and finding again a regime of saturation. 
        
        For intermediate mass stars, between 3\,\Msun\  and 10\,\Msun\  the X-ray luminosity undergoes a sharp decrease, and it is hard to find precise values in literature, considering both the large scattering of the observations and the different physical processes behind the X emission. For these reasons, we adopted the constant value of $ L_{\mathrm{ XUV}} \approx 10^{-6}  L_{\mathrm{bol}} $ as upper bound of common literature values, in this mass range (see, e.g., \cite{Flaccomio03a:XUV, Flaccomio03b:XUV, Naze09:X, Gorti09:X}).
        
        For the EUV component of the high-energy flux, we have used the equation of \cite{ForcadaMicela11:XUV} (without error bars):
        \begin{equation}   
            \log(L_{\mathrm{EUV}}/\si{erg\ \second^{-1}}) = 4.8 + 0.86 \cdot \log(L_{\mathrm X}/\si{erg\ \second^{-1}}) 
        \end{equation}
        which relates $ L_{\mathrm{EUV}} $ to $ L_{\mathrm X} $ and assumes that X and EUV luminosities have a uniform development in time. 
        If the X-ray emission literature is uncertain on the $ L_X $ of intermediate-to-heavy stars, the EUV counterpart is even more obscure, because of the sheer lack of available data, due to the severe absorption of these wavelengths in the interstellar space. We then extended the relation of \cite{ForcadaMicela11:XUV}, that was originally calibrated on M to late-F type stars, to stars up to 3\,\Msun, at any time before the WD stage, to have a rough best estimate of the ultraviolet component.
    
        The photoevaporation can continue even after the death of the stars, as they become white dwarfs, and depending on the orbital distance of the planet it can actually have a significant impact on its atmosphere, given the extremely high temperatures of the WD surfaces, right after their formation \citep{Schreiber19:WD}. We then deemed worth including the contribution of these compact objects in our high-energy computation for the evaporation. Considering the wide range of possible temperatures of the newborn WDs, up to $ \sim 10^5 \si{\kelvin}$, the average large distance of the planet from the inner WD and the lack of published stellar models in these high-energy part of the spectrum, we approximate the white dwarf as a blackbody. Thus, the code integrates the blackbody specific flux, or radiance, $ B_{\lambda} (T) $, on the wavelength range $ 10 - 912 \si{\angstrom} $ and on the star's surface area, to obtain the overall XUV ionising luminosity. In this way a simple temperature-dependent relation is computed for every possible WD, evolving in time as they cool down, regardless of the chemical composition. This approximation represents an overestimate of a real white dwarf's high-energy flux: real WDs have H/He-rich atmospheres that re-absorb part of this ionising radiation. However, this approximation is sufficient to serve the purposes of this work, also considering that we deal with generally wide-orbits CBPs, after stars have already expelled their shells.

    \subsubsection{Stellar Winds}
        \label{subsubsec:stellar_winds}

        Stars considered in this study (i.e., stars with an initial mass range of $0.95\, M_{\odot} \lesssim M_{\rm ZAMS} \lesssim 10\,M_{\odot}$) lose a significant fraction of their mass in their evolved stage. Such stars can lose up to 40$\%$ of their initial mass via so-called dust driven winds during their asymptotic giant branch phase. At this evolutionary phase, the star rapidly expands and their outer layers cool down sufficiently for dust formation. The momentum imparted onto the dust particles by radiation will then drive the mass loss. This mechanism can result in mass loss rates as high as $\dot{M} \sim 10^{-7}$-$10^{-4}\,M_{\odot}yr^{-1}$ (see e.g., \citealt{Hofner2015}). Furthermore, the evolution of  O/B stars (roughly corresponds to initial masses $M_{\rm ZAMS} \gtrsim 8\,M_{\odot}$) is also significantly affected by line-driven winds (e.g.,\citealt{Puls2008}). Such massive stars, therefore experience significant mass loss from the start of the main sequence phase.

        The mass loss rates of stellar winds and their effects on stellar evolution are determined by \verb|SeBa|, while the effects on the orbit of the triple are determined by \texttt{TRES}. How the properties of stars change due to stellar winds in \verb|SeBa| is explained in \citet{Toonen12:DWD} in detail, which is based on the formalism of \citet{Hurley02}. The stellar wind prescriptions used in \verb|SeBa|, which provide an estimate on the mass loss rates based on the properties of the star, are introduced in detail in \citet{Toonen12:DWD}, but we also provide a brief summary below.
     
        If the star is on the main sequence or a Hertzsprung gap star and has a luminosity larger than $L>4000\,L_{\odot}$ and an effective temperature \teff $>$ 8000\,K, we assume that line-driven winds are efficient and we calculate the mass loss rates according to \citet{Vink2001}, if \teff $\leq 50 kK$ or according to \citet{Nieuwenhuijzen1990}, if \teff $>$ 50 kK. If the effective temperature of the star is below \teff = 8000$\,K$, we assume line-driven winds are no longer efficient and instead dust-driven winds dominate. In this case, we calculate the mass loss rates according to \citet{Nieuwenhuijzen1990} and \citet{Reimers1977} and take the maximum of these two values. If the star, however, is on the asymptotic giant branch, besides the mass loss rates of \citet{Nieuwenhuijzen1990} and \citet{Reimers1977}, we also calculate \citet{Vassiliadis1993} rates to account for the so-called "superwind" phase of thermally pulsating asymptotic giant branch stars. The final mass loss rate is the maximum value given by these three different prescriptions.

        In order to compute the change in the orbit due to stellar winds, we assume stellar winds are fast, spherically symmetric. We furthermore neglect accretion by companions. In that case the inner and the outer orbit of the triple widens in an adiabatic fashion as (see e.g., \citealt{Soberman1997}):
        \begin{equation}
            \dot{a}_{\rm in,wind} = \left( \frac{a_{\rm final}}{a_{\rm init}}\right)_{\rm in} = \frac{M_{\rm, 1,init} + M_{\rm 2,init}}{M_{\rm 1,final} + M_{\rm 2,final}},
        \end{equation}
        and
        \begin{equation}
            \dot{a}_{\rm out,wind} = \left( \frac{a_{\rm final}}{a_{\rm init}}\right)_{\rm out} = \frac{M_{\rm 1,init} + M_{\rm 2,init} + M_{\rm planet}}{M_{\rm 1,final} + M_{\rm 2,final} + M_{\rm planet}},
        \end{equation}
        where subscripts $init$ and $final$ refer to properties before and after the stellar winds carried mass away from the inner binary in a given timestep.
        The assumption of adiabatic expansion caused by stellar winds is justified by two arguments. Firstly, the terminal velocities of stellar winds are typically much larger than the stellar orbital velocities, meaning the wind does not interact directly with the orbit \citep[see e.g.][]{Vink2001}. Secondly, the  timescale of the orbital change is typically much longer than the mass loss rate timescale, and therefore the mass loss does not vary within a period (as long as the orbit is not wider than a few thousand au, see discussion in \secname~\ref{sec:massloss}).
        Neglecting wind accretion is justified for line-driven winds, however, the dust-driven winds of asymptotic giant branch stars/supergiants can have velocities as low as $v\approx5$-$30\,\rm{kms^{-1}}$. In the latter case, the accretion efficiency could be as high as $\sim 50$ per cent (\citealt{Hofner2015}).
        We assume that the eccentricity remains unchanged by stellar winds (\citealt{Huang1956A}, \citealt{Huang1963}).

    \subsubsection{The inclusion of SSOs mass-radius relation}
    
        In order to give a spatial dimension to the SSOs under analysis, we implemented an equation to relate the mass of a SSO (i.e., the input to \texttt{TRES}) to its radius. In fact, the radius of a planet, and thus its density with a given mass, affects several processes during its dynamical evolution in a triple, e.g.: 
        both tidal interactions and the photoevaporation are affected, not to mention the moment of inertia and all quantities coupled to that.  
        In the literature there are different kinds of works that use simple analytic relations to determine the size of a SSO. Some have a more theoretical approach and exploit the polytropes \citep[e.g.,][]{Chabrier+2009:radii, GuillotGautier2015:giants} and others take an empirical approach \citep[e.g.,][]{Chen&Kipping2017:radii, Bashi+2017:radii, Thorngren2019:radii}. Eventually, we adopted the simple analytic expression of \cite{Chen&Kipping2017:radii}, for their rigorous statistical analysis applied on a large sample of empirical data, spanning the widest mass range  (from SSOs of \SI{\sim e-3}{M_\oplus} to \SI{0.87}{M_\odot} stars). The authors compiled a table of 316 objects in this range, with well-defined mass and radius values, and employed hierarchical Bayesian modelling to obtain a probabilistic broken power-law. They have created a \texttt{Python} package to easily allow the use of their results, called \texttt{Forecaster}. 
        For our purposes, we only implemented in \texttt{SeBa} the deterministic power-laws in the form:
        \begin{equation*}
            R = C \cdot M^p ,
        \end{equation*}
        with the radius R that is a function of the mass M to the power p and coefficient C, where these two coefficients vary in mass sub-intervals as prescribed by \cite{Chen&Kipping2017:radii}.

    \subsubsection{SSOs spin velocity}        

        The last piece of knowledge needed to properly initialise our giant planets was the rotational velocity, or spin, around their axis at ZAMS. 
        Planets acquire rotation during their formation, when the accreting matter brings its angular momentum to the object; in principle, then, knowing the formation history of a planet would provide its initial spin velocity. 
        Unfortunately, planet formation processes are still unclear and have major uncertainties; anyway, we did not simulate the earliest phases of stellar and planetary formation, thus we needed a realistic value of the spin at ZAMS, depending only on the mass of the object and regardless of its specific accretion history.

        We decided to use the estimate of \cite{Bryan+2018:spin}, who compiled a sample of young substellar objects up to 20\,\Mj, with rotation rates constrained by observations (in part by emission line broadening).
        They found that the spin velocity of the sample is approximately an order of magnitude lower than the break-up velocity, which corresponds to the maximum possible rotation rate to remain gravitationally bound to a body, depending on its mass and radius. 
        Without a process dissipating the angular momentum, the gaseous planets would spin up to near the break-up velocity, whereas we observe terminal speeds several times lower that this limit. 
        The main mechanism responsible for this dissipation is thought to be the magneto-hydrodynamical coupling between the gas giant and its circumplanetary accretion disk, expelling angular momentum away \cite{Batygin2018:spin, Bryan+2020:spin, Dittmann2021:spin}.

        Another interesting remark of \cite{Bryan+2018:spin} concerns the little evolution of the terminal spin rate. Their sample of young SSOs is aged between 2-200\,Myr but they cannot find any time-dependence of the spin: considering that Jupiter's and Saturn's present-day velocities are compatible with the sample mean, the authors infer that the rotation of these objects is set in the earliest phases of formation and then does not evolve significantly for billion of years.
        For the reasons above, we adopted as the rotation velocity of the SSOs at the ZAMS their best-fit mean value: 
        \[ v_\mathrm{ZAMS} = 0.126 \cdot v_\mathrm{break-up}  \]
        All SSO are assumed to be spin-orbit aligned and the spin evolves as a scalar quantity.

\section{Population synthesis}
    \label{sec:popsyn}
    
    For the long-term evolution of circumbinary giant planets, we have defined a set of priors for the inner stellar binary, and two sets of priors for the planetary companion which we run separately.
    Details on the set-up for the orbital features of both inner binaries and planets are described below, as well as with the specification on the \texttt{TRES} simulations characteristics and population setup.

    \subsection{Inner binary priors}
        \label{sec:binpriors}
    
        To initialise the inner binaries at ZAMS we randomly sample from the following prior distributions, for the inner binary separation $a_{\rm in}$, the mass ratio $q = M_2/M_1$, the primary star initial mass $M_1$, and the binary eccentricity $e_{\rm in}$:  
        \begin{itemize}
        \setlength\itemsep{0.3em}
            \item[-] $a_{\rm in}$: $\log{}_{10}$-uniform distribution 
            $\log{\mathcal{U}_a}$ for $ a \in $ (15\,\Rsun; 2200\,\Rsun)  \citep{Abt1983, Kouwenhoven2007}, \citep{DucheneKraus2013};
            \item[-] $M_1$: Kroupa's mass function  $n(M_1) \sim M_1^{-2.35}$ , for $ M_1 \in  ( 0.95, 10)\,$\Msun \ \citep{Kroupa1993};
            \item[-] $q$:
            Uniform distribution $\mathcal{U}_q$ for $ q \in (0, 1] $ (e.g., \citealt{Raghavan2010, Sana2012, MoeDistefano2017, DucheneKraus2013}) with a lower limit of 0.95\,\Msun \ on $M_2$;
            \item[-] $e_{\rm in}$: Thermal distribution  $ f(e) = 2e $ , for $ e \in (0, 0.95) $ \citep{Heggie1975}.
        \end{itemize}
        
        This set of priors was chosen specifically to generate a population of binary stars which could reproduce the observed Milky Way DWD distributions \citep{Toonen12:DWD, Toonen2017}. 
        The upper limit on $ a_{\rm in} $ is influenced by LISA detection limits.
        The mass boundaries are chosen to maximise our computational efforts specifically on those binary stars evolving on timescales comparable with one Hubble time and thus, be able to leave white dwarf remnants at their death, for the most part. 
        The initial $q$ of the simulated binaries must always satisfy the given mass boundaries (see the points above), implying that the actual minimum in our simulations is $q_\mathrm{min} = 0.095$.

    \subsubsection{Common-envelope evolution}
        \label{sec:CE}
        
        Common-envelope (CE) evolution is an important process in binary evolution \citep[see e.g.,][]{Ivanova2013, Iaconi19, Ivanova20}, thought to be responsible for the formation of the majority of compact star systems \citep[e.g.,][]{Izzard12}. During this process the two stars share a CE (hence the name) that is not in co-rotation with the orbital motion and as a result the stars spiral closer to each other. The end result of a CE-phase is either a merger of the two stars or a successful ejection of the CE material leaving behind a compact binary.
        
        Despite the importance of the phase, and the large effort of the community the CE-phase is poorly understood and constrained. Major discussions concern the nature of the energy source that is responsible for unbinding the CE, such as orbital energy \citep{Paczynski1976,Webbink1984}, recombination energy \citep{Ivanova15,Nandez15, Grichener18, Kramer20, Reichardt20, Lau22b}, ionisation energy \citep{Han94,Han95,Sand20}, radiation energy \citep{Ivanova18, Lau22}, accretion \citep{Chamandy18}, convection \citep{Sebach17,Ivanova18, Wilson19, Wilson22}, jets \citep{Shiber18,Shiber19, Lopez22}, and dust \citep{Glanz18, Iaconi20}. Purely hydrodynamical simulations typically do not unbind the envelope, although some success has been achieved recently by including other energy sources than orbital energy \citep{Ivanova2016, Law2020}. 
        
        In addition to hydrodynamical studies, a few observational constraints exist. These arise from studies of individual post-CE systems whose evolution is modelled backwards in time \citep{Nel00, Sluys2006, Zorotovic10, Zorotovic22} or by comparing the observed population demographics to the theoretical models (so called population synthesis models) \citep{Toonen&Nelemans13:CE, Camacho2014}. Specifically for double white dwarfs, this has lead to the following insights into their formation \citep{Nel00, Nel01,Sluys2006}: 1) the first phase of mass transfer leads to a moderate widening of the orbit; 2) the second phase of mass transfer (in which the second white dwarf is formed) leads to a strong shrinkage of the orbit. We follow the suggestion of \cite{Nel01} to model the CE-phase with the classical $\alpha$-prescription with parameters $\alpha\lambda=2$ when the companion is a compact object or when the CE is triggered by a tidal instability \citep[rather than dynamically unstable Roche lobe overflow][]{Darwin1879}, and the $\gamma$-prescription with parameter $\gamma=1.75$ otherwise.

    \subsection{Circumbinary planets priors}
        \label{sec:CBPpriors}
    
        To account for different outcomes in the final population, which can arise as consequence of the initial assumptions, we defined two different sets of initial orbital parameters distributions for our simulations.  
        Therefore, we modelled two different populations to study the impact of the planetary orbital priors on the final survival and ejection rates. We note that for both synthetic populations we kept the binary priors (\secname~\ref{sec:binpriors}) and the CE assumptions (\secname~\ref{sec:CE}) constant. The analysis on the consequence of employing different binary evolution models on the long-term evolution of CBPs will be presented in an upcoming study.
        As previously mentioned, the low statistics of CBPs does not allow us to robustly trace the typical properties of their orbital parameters, for such we refer to those observed for single-star planetary systems. In particular, for the first population, which we labelled {\it Population A} (Pop. A), we employed 
        the distributions reported in \cite{Danielski19:CBP} and references there-in. Given our interest for the $Magrathea$ planets, in the framework of the science development of the LISA mission, we adopted their optimistic scenario (B1) distributions for both the semi-major axis $a_{\rm out}$, planetary mass $M_P$, and orbital inclination $i_P$.
        In addition, we included the eccentricity $e_{\rm out}$ distribution by \citep{Bowler2020}, and used in \cite{Katz2022} for studying the CBP detection efficiency by LISA.
        In details, the distributions used to draw initial parameters for Pop. A are the following:
        
        \vspace{-0.4em}
        \begin{itemize}
        \setlength\itemsep{0.3em}
            \item[-] $a_{\rm out}$:\ \ $\log{}_{10}$ Uniform distribution: $\log{\mathcal{U}_a}$ (0.17\,au; 200\,au);
            \item[-]$i_P$:\ \  Uniform distribution $\mathcal{U}_{\cos{(i)}}$ ($-1; 1$);
            \item[-]$M_P$:\ \  Uniform distribution: $\mathcal{U}_M$(0.2 \Mj;  16 \Mj);
            \item[-]$e_{\rm out}$:\ \  Beta distribution $f(e|\alpha, \beta) = \frac{\Gamma(\alpha + \beta)}{\Gamma(\alpha)\Gamma(\beta)} e_2^{\alpha-1}(1-e)^{\beta -1}$, in the range (0; 0.95) where $\Gamma$ is the gamma function, $\alpha$ = 30, and $\beta$ = 200.
        \end{itemize}
        
        The planetary mass range was chosen to only span the giant planets mass spectrum, to focus on those planetary companions that have the chance to be detected by LISA \citep{Danielski19:CBP,Katz2022}. The upper mass limit was chosen based on \cite{Spiegel2011}, instead of the hard deuterium limit.
        The orbital separation lower boundary $a = 0.17$\,au, between the planet and the centre of mass of the binary, was defined based on the stability limit by \cite{Holman&Wiegert99:stability}, when computed for a binary with the smallest circular orbit in the binary population (see \secname~\ref{sec:binpriors}) and equal stellar mass M$_1$ = M$_2$, meaning $ \mu = 0.5 $ in Eq.~\ref{eq:stability}.
        
        Concerning the second population, i.e., {\it Population B} (Pop. B), we adopted uniform distributions for all parameters to avoid population biases that might be caused by priors based on single-star planetary systems observational evidence. Nonetheless, the range spanned by each parameter is the same used for Pop. A.

        \begin{figure*}[tb]     
            \centering
            \includegraphics[width=0.8\linewidth]{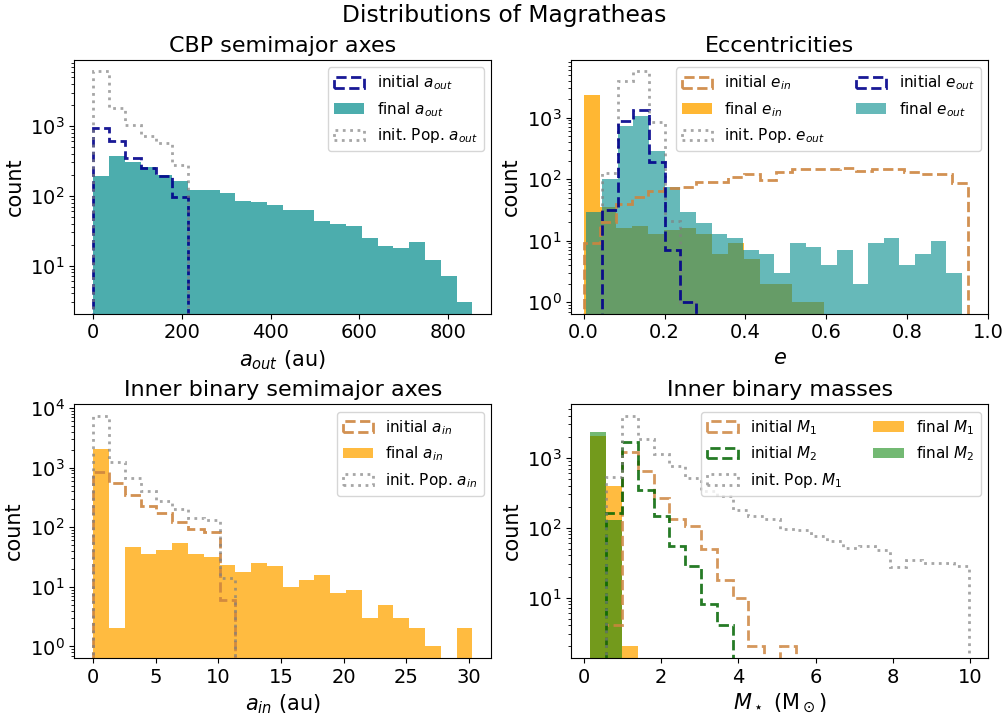}
            \caption{{\it Population A:} overview of the distributions of Magrathea systems in the parameter space. Solid histograms represent the Magrathea parameters at one Hubble time, while the dashed lines show their initial distributions. The "out" subscript denotes the planetary parameters (blue distributions). The dotted grey lines, instead, show the initial distributions for the whole population, not restricted to the Magratheas.} 
            \label{fig:quad_magratheas_lit}
        \end{figure*}

    \subsection{Simulations setup}
        \label{sec:simulations}
        
        Once chosen the initial distributions for our stars and planets, we generated a statistically representative sample of our two populations, simulating eventually 10500 systems for each population.
        Each new system is generated by independently sampling two stars and one SSO based on their respective priors.  However, since \texttt{TRES} code can only simulate secularly-stable systems, the random draws of initial parameters are checked against stability at zero age. The system is discarded if its random initialisation is unstable and a new extraction is performed. 
        We imposed an upper limit of 2 hours to the CPU time allowed for each evolution phase, to avoid cluttering our population synthesis with few systems requiring an anomalously large amount of computing time. These computationally-heavy simulations mostly corresponded to systems lying close to the triple's stability limit, where the secular approximation timestep was pushed to extremely small values, due to fast-evolving three-body dynamics. A quick overview of these systems is given in \secname ~\ref{sec:cpu_limited}.
        
        CPU time limits aside, we stopped the simulation of those systems in which any of the components merge, becomes unbound, initiate a phase of stable mass transfer or if the triple becomes dynamically unstable, in general. 
        It is currently challenging to predict the outcome of a stable phase of mass transfer in the inner binary of a triple system, as the orbit tends to be eccentric due to three-body interactions (see also the \secname ~\ref{sec:res_rlofed} ). 
        However, systems undergoing CE phases were not stopped, which allowed us to obtain tight inner binaries and remnants.
        We set the maximum simulation time as one Hubble Time, defined as \SI{13.5}{Gyr}, after the start from the ZAMS. Thus, if a system had not stopped earlier by any of the conditions outlined above, the code would stop the simulation at this moment.

\section{Results}
    \label{sec:results}
    
    To analyse the data, we subdivided the synthetic populations (A and B, see \secname~\ref{sec:popsyn}) into categories, based on their final fate. 
    The main categories we identified are the following:
    \begin{description}
        \item [--] {\it Magratheas:} circumbinary planets survived to the full inner binary evolution from ZAMS up to \SI{13.5}{\giga yr} (i.e., a Hubble time), with both stars turned into white dwarfs;
        \item [--] {\it Collided:} systems whose evolution stopped once the circumbinary planet's orbit intersects with the inner binary's orbit. 
        The subsequent fate of these planets cannot be simulated with a secular code;
        \item [--] {\it Destabilised:} systems that become dynamically unstable, either due to the three-body dynamics, to disruptive evolution events (e.g., supernovae), or due to adiabatic stellar winds . 
        \item [--] {\it Merged:} systems whose inner binary stars merge before one Hubble time;
        \item [--] {\it Stable-MT:} systems whose inner binary initiate a phase of stable mass transfer, which was also a stopping condition in our simulation.
    \end{description}

    The percentages of simulated systems falling into each one of the described category, for both populations, are reported in \tablename~\ref{tab:percentages}. 
    To avoid cluttering the body of the paper, a complete collection of the plots for all these main categories can be found in the Appendix. 
    For completeness, we mention the presence of two other secondary categories: the \emph{CPU-limited} systems, introduced in \secname~\ref{sec:simulations} and discussed in \secname~\ref{sec:cpu_limited}, and the \emph{Ordinaries}, a collection of the remaining systems which did not belong to any other category of interest of ours and that basically consisted of half-evolved systems with long-lived lighter mass stars (amounting to around 11\% of the total sample).

    \begin{table}[tbh]         
        
        \centering
        \begin{tabular}{l  r  r }
            \toprule
             & Population A & Population B \\
            \midrule
            Magrathea  &   23.21 \%    &   32.10 \%    \\
            Collided    &   3.18 \%     &   2.11 \%     \\
            Destabilised    &   0.26 \%    &   0.17 \%    \\
            Merged      &   31.70 \%    &   35.10 \%    \\
            Stable-MT   &   16.94 \%    &   17.08 \%    \\
            CPU-limited   &   12.01 \%    &   2.47 \%    \\
            Ordinaries   &   10.70 \%    &   10.71 \%    \\
            \bottomrule
            \end{tabular}
        \caption{Percentages of possible evolution outcomes computed over the total size of Pop. A and B (both with a sample size of 10500). The characteristics of the two populations are described in \secname~\ref{sec:CBPpriors}.}
        \label{tab:percentages} 
    \end{table}

    \subsection{Magrathea systems}
        \label{sec:res_magratheas}
    
        \begin{figure}[t]          
            \centering
            \includegraphics[trim=0.0cm 0.1 0.1cm 0.5 0cm,clip,width=1\linewidth ]{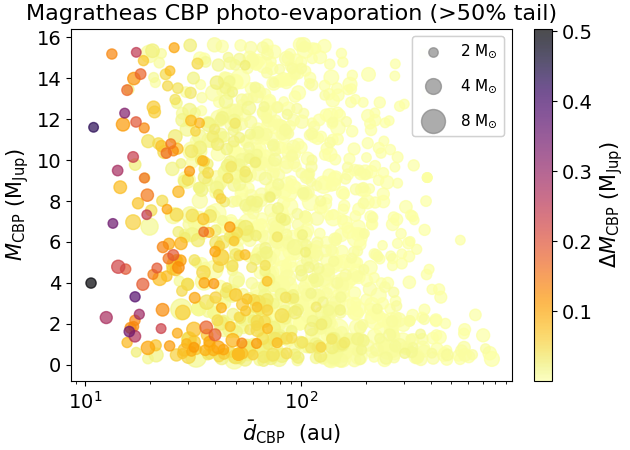}
            \caption{{\it Population A:} Atmospheric mass loss endured by Magrathea gas giants, scattered in the mass-distance parameter space. The x-axis corresponds to the final time-averaged orbital distance of the CBPs. We show the planets which lost more mass than the 50° percentile of the category. The colour corresponds to the amount of mass lost (as shown in the colour bar). The size of the markers is proportional to the progenitor binary mass, as illustrated in the legend. } 
            \label{fig:photoevap_magratheas_popA}
        \end{figure}

        \begin{figure}[tb]          
             \centering
            \includegraphics[width=\linewidth ]{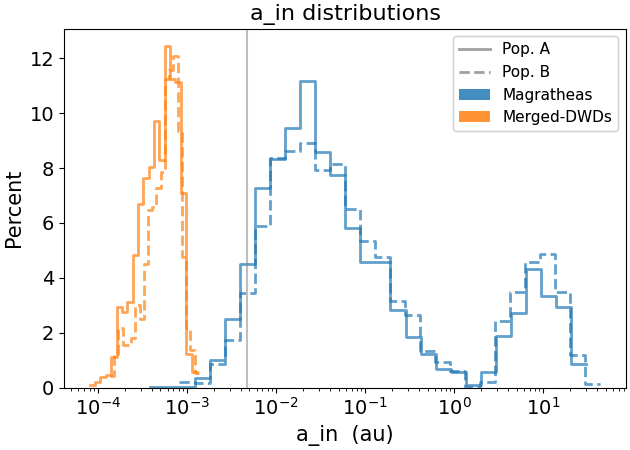}
            \caption{ Comparison of the inner semi-major axis $a_{\rm in}$ distributions for the binaries in Magrathea systems (blue) and the merged-DWDs (orange, $a_{\rm in}$ refers to the timestamp just before merging) of population A and B (line and dashes, respectively). The grey line denotes a separation of  1\Rsun.} 
            \label{fig:binsep_magra_dwdmerged}
        \end{figure}

        The main interest of our analysis are \emph{Magrathea} planets. Our results show that a significant fraction i.e., between the 23\% -- 32\% of the generated triples survive to become systems with a circumbinary planet orbiting a DWD (\tablename~\ref{tab:percentages}).
        
        {\it Population A:} $\sim$ 23\% of planets among the total systems become Magratheas by one Hubble Time. An overview of the ensemble properties of this population is given in \figurename~\ref{fig:quad_magratheas_lit}.
        The giant planets have final orbital parameters which overall preserved their initial distributions shape, but with a larger spread in the final values, particularly for what it concerns their semi-major axes $a_{\rm out}$  (Fig. \ref{fig:quad_magratheas_lit}, top left), and less evident for the eccentricity $e_{\rm out}$ (Fig. \ref{fig:quad_magratheas_lit}, top right).
        The majority of the host stars resulted in tight binaries with circularised orbits (\figurename~\ref{fig:quad_magratheas_lit}, top right panel).
        The stellar progenitors of Magrathea systems are biased towards the lighter masses, peaking around $1 - 2$ \Msun \ (Fig. \ref{fig:quad_magratheas_lit}, bottom right).
        The photo-evaporation eroded on average 0.13\% of the gas giants starting mass, with the most extreme case losing almost 0.5\,\Mj\ during their life. The Magrathea percentage, although not large in absolute terms, is one of the highest among our categories, beaten only by Merged and Ordinaries. We show the impact of photoevaporation as function of the planet's mass and orbital distance in \figurename~\ref{fig:photoevap_magratheas_popA}, where each point of the scatter plot is a Magrathea CBP, coloured according to the endured mass-loss.
        The time-averaged orbital distance, on the x axis of \figurename~\ref{fig:photoevap_magratheas_popA} was calculated as described in \secname~\ref{sec:evap}.
        \figurename~\ref{fig:photoevap_magratheas_popA} displays a strong correlation between the evaporated mass and $ \Bar{d} $, as expected, and a weaker correlation with the initial planetary mass. We note, moreover, that the heaviest photoevaporation is not caused by the largest progenitors.
    
        {\it Population B :} $\sim$ 32\% of total systems become Magratheas by one Hubble Time, showing that the survival rate of planets decreases for the shortest orbital distances, on average. 
        In fact, a wide-orbit planet suffers lesser effects of stellar evolution than those on closer orbits. However, we remind the reader that we are assuming both adiabatic stellar winds (see \secname~\ref{sec:massloss}) and an isolated gravitational environment i.e., we are excluding any external contribution from nearby systems, that could likely result in stripping the widest orbits' planets.
        In comparison with Pop. A,  Pop B presents a significant difference in the planetary eccentricity distribution, which is mostly a consequence of the initial distributions (\figurename~\ref{fig:quad_magratheas_B}, top right). This may imply that the CBP eccentricity is not by itself a decisive parameter to determine the survival around binaries. In absolute terms, the eccentricities have not been dramatically altered  during the systems evolution, so that the final distribution shape follows the initial one. 
        Also in this population, the planetary semi-major axes cover a large range of values up to $a_{\rm out} = 1100 $\,au, slightly higher than Pop A. 
        Photo-evaporation generated a mean loss of just 0.02\% of the initial mass, but with a maximum evaporation of 0.6\,\Mj\ similarly to Pop. A. 
        The inner binaries show overall properties similar to those of Pop. A.

        \begin{figure}[t]         
            \centering
            \includegraphics[width=\linewidth ]{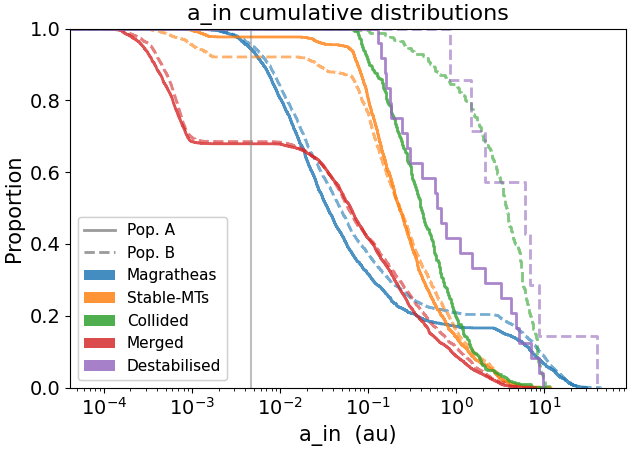}
            \caption{ Complementary cumulative distributions of the final inner binary semi-major axes $a_{\rm in}$, for all categories (colour coded in the legend), and for both Pop. A (solid lines) and B (dashed lines). Merged DWDs all resides in orbits smaller than 1\Rsun \ (marked by the vertical grey line) and few of the Magrathea hosts too.
            These parameters refer to the last valid simulation step in secular approximation.
            }
            \label{fig:cumul_a_in}
        \end{figure}

\begin{figure}[t]        
            \centering
            \includegraphics[width=0.98\columnwidth ]{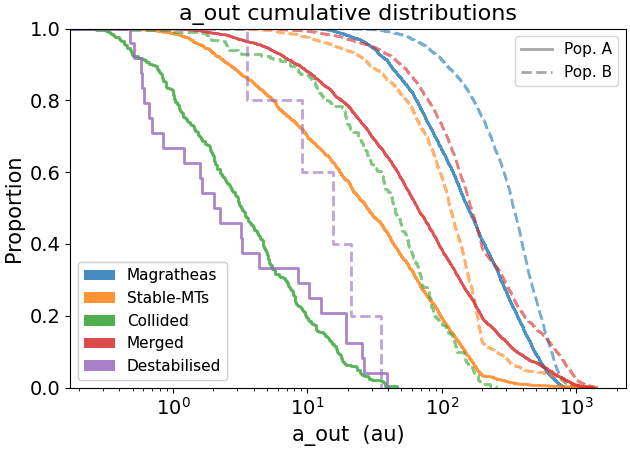}
    
            \caption{Complementary cumulative distributions of final semi-major axes $a_{\rm out}$ of CBPs for all categories (colour coded in the legend), for both Pop. A (solid lines) and B (dashed lines). 
            These parameters refer to the last valid simulation step in secular approximation.
            }
            \label{fig:cumul_a_out_all}
        \end{figure}

\begin{figure}[t]          
            \centering      
            \includegraphics[width=0.95\linewidth ]{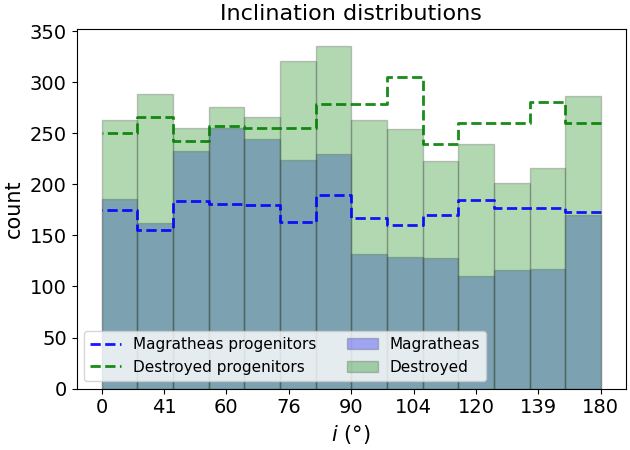}
            \caption{{\it Population A:} relative inclination distributions for Magrathea and planets in Destroyed systems (which include Merged, Collided, and Destabilised). The Population was initialised with an inclination distribution uniform in $ \cos{(i)} $; this plot illustrates how at the end of life the systems seem to prefer mildly-inclined prograde orbits, especially surviving Magratheas. }
            \label{fig:PopA_hist_inclination}
        \end{figure}

        \textit{Both populations:} No appreciable selection effect for the planetary masses is noted for this class of systems.
        Binary stars have mass ratios peaking at the unit value. 
        The orbital separations of the binary stars range from a few $10^{-3}$\,au to a few tens of au, with a gap around a few au (\figurename~\ref{fig:binsep_magra_dwdmerged} and \figurename~\ref{fig:cumul_a_in}). This gap, as also recently confirmed with Gaia observations \citep{Korol22}, separates the post-mass-transfer binaries (with smaller circularised orbits) from the binaries that are too wide to experience a mass transfer phase in any point in their evolution.
        \\
        From the top two plots of \figurename~\ref{fig:semiaxes_ratios}  it can be seen, again (see also the eccentricity panels in \figurename~\ref{fig:quad_magratheas_lit} and \figurename~\ref{fig:quad_magratheas_B}), how the Magrathea planets at one Hubble Time are in the most part orbiting circularised binaries: based on the stability equation described in \secname~\ref{subsec:stability}, in fact, for null eccentricity and equal-mass binaries (which is the case for most Magrathea hosts) the critical semi-major axis ratio assumes the particular value of $ a_{crit} = 2.39 $, where most of the CBPs are found. The giant planets themselves, however, are scattered across several order of magnitudes multiples of the critical semi-major axis, not showing preferred stability regions for their orbits where they could strongly pile up (see top panels of \figurename~\ref{fig:semiaxes_ratios}). 
        When comparing the cumulative distribution of final CBPs semi-major axes across the different categories (\figurename~\ref{fig:cumul_a_out_all}), we see that Magrathea planets reside on the largest orbits of all categories, except for a small tail of the Merged category. 
        The stellar hosts, on the other hand, have rather tight orbits, rivalled only by some of the tightest merging DWDs (see \figurename~\ref{fig:cumul_a_in}). This creates wide and stable hierarchical triples that can survive until a Hubble time. \\
        Finally, we found an asymmetry in the final distribution of relative inclinations of the Magrathea planets: they generally preferred prograde orbits and have a broad peak at around \ang{60} from the stellar orbit plane. In \figurename~\ref{fig:PopA_hist_inclination} for Pop. A and \figurename~\ref{fig:PopB_hist_inclination} for Pop. B, it is visible how the final distribution show an over-abundance prograde planets which amounts to $+12.6\%$ and $+7.8\%$, respectively.
        In those same figures, we defined as the \emph{Destroyed} systems those ending with a ``catastrophic'' event for the triple, consisting of the Merged, Collided and Destabilised systems collected together for convenience. The Destroyed too registered a gain in the prograde orbits at the end of the evolution, but a factor 2.3 and 4.6 times less than Magratheas, respectively in Pop. A and B, and closer to the 90° inclination.
        This bias was not present in the initial distribution which was set to uniform in $\cos(i)$. 
        This is consistent with recent insights in the Lidov-Kozai mechanism for systems outside the test-particle regime \citep{Anderson2017, Lei19, Hamers21}. In this model the maximum eccentricity is not reached at a mutual inclination of 90°, but at slightly higher values. As those systems are more likely to interact (e.g., merge), we expect the Magrathea planets to prefer prograde orbits.

    \subsection{Collided}
        \label{sec:res_collided}
    
        The Collided systems, in which the CBP orbit intersects the inner binary orbit, belong to a category with quite distinctive parameters; they are a small but non-negligible fraction. 
        The most characterising feature is the very high eccentricity of all the giant planets in this category, on average higher than for all the other categories, in both populations (see the green lines in \figurename~\ref{fig:cumul_e_out}). 
        This category is in some aspects similar to the Destabilised (see \secname~\ref{sec:res_destabilised}), since the configuration of the triple is severely altered and the assumptions for a hierarchical stability are violated at the ending time. Yet, while the Destabilised category simply accounts for any triple that become unstable at some point in its evolution, the Collided include specifically those triples where the CBP acquired an orbit eccentric enough to intersect the inner binary orbit, and thus likely go through phases of fast dynamical re-shaping. 
        The eccentricity of these CBPs orbits reaches high values before they can result unstable, based on the stability criterion described in \secname~\ref{subsec:stability}. Eventually, for their peculiar properties and their abundance, we present them separately from the Destabilised category. 
        The final parameters of the Collided category are to be carefully interpreted as  they refer to the last simulation step, before the stop due to violation of the secular evolution assumptions. The label of such category should indeed not to be confused with a real physical planetary collision scenario.
        
        {\it Population A:} around 3.2\%. These planets have an entirely different final eccentricity distribution than its initial one, with no overlap between the two, and with the latter having always $ e_{\rm out}  > 0.5 $, as illustrated in the top right panel of \figurename~\ref{fig:quad_grilled}.
        Their semi-major axes are relatively small, extending at most to few tens of au, yet planets are piling up close to the stability limit (see \figurename~\ref{fig:semiaxes_ratios}, second  row left). 
        Stellar hosts as well show higher values of eccentricity than Magrathea systems, and their orbit does not circularise. The orbital separation in the inner binaries spans a shorter interval, approximately $a_{\rm in}$ = 0.07 -- 10\,au (see \figurename~\ref{fig:cumul_a_in}).
        The stellar types of this population mainly consists of Main Sequence stars (see central panel of \figurename~\ref{fig:startypes_A}), confirming again the short lifespan.

        {\it Population B:} 2.1\% of the total sample. For this population we see that the progenitors' planetary eccentricities are skewed towards the highest values possible, peaking right before $e_{\rm out}$ =  0.95, as it is illustrated in the bottom panel of \figurename~\ref{fig:quad_grilled}. However, the final distribution is devoid of low-eccentricity planets and, interesting enough, it covers eccentricities $e_{\rm out} > 0.5 $, similarly to Pop. A, notwithstanding the completely different initial distributions.
        The planetary semi-major axes in this population are spread up to $a_{\rm out}$ = \SI{200}{au}, and those of the inner binaries up to $a_{\rm in} = $ \SI{10}{au}, showing larger orbits than in Pop. A (see \figurename~\ref{fig:quad_grilled}). The range of stellar mass covered is the same as the one of Pop. A. In \figurename~\ref{fig:cumul_a_in} it is remarkable how the inner binary separation of Pop. B covers all the range up to the upper boundary of \SI{\sim 10}{au}  in a more homogeneous way. 
        We noticed, as well, that in this population there seems to be a dichotomy by age of the Collided: the majority has ages which range from 10 to $ 10^4 $ \,Myr, while another small part of the sample is younger covering ages down to $ 10^{ -4}$\,Myr, which is indicative of systems born already on the brink of instability and that could realistically never make it to ZAMS. The statistics of the smaller portion is though not sufficient to assess a departure of its features within the Collided category.
        When the evolution is interrupted, the stars are for a quarter post-MS stars (see central panel of \figurename~\ref{fig:startypes_B}), different than in population A. 
    
        \textit{Both populations:} atmospheric mass loss from the planets is negligible, likely because of the limited lifetime of these systems before the orbital collision. The maximum single mass evaporated is only $ 10^{-3} $\,\Mj. 
        On the other hand, the lifetime is long enough to alter the eccentricity but not the semi-major axes of planets and the stars.

    \subsection{Destabilised}
        \label{sec:res_destabilised}
    
        The Destabilised systems are a tiny fraction of the total, $\sim$ 0.2\% of the simulated triples for both populations, and they touch extremes of the parameter space. This category includes the system that became unstable due to chaotic three-body dynamics, or which are disintegrated by a violent supernovae kick. 
        \texttt{TRES} discards automatically systems that are already unstable at their random initialisation, at $t=0$, so the Destabilised category comprehends only those triples which became unstable after the ZAMS. We found these systems to be characterised by initial orbital parameters values near the stability limit.
        
\begin{figure*}[tb]       
            \centering
            \includegraphics[width=\linewidth ]{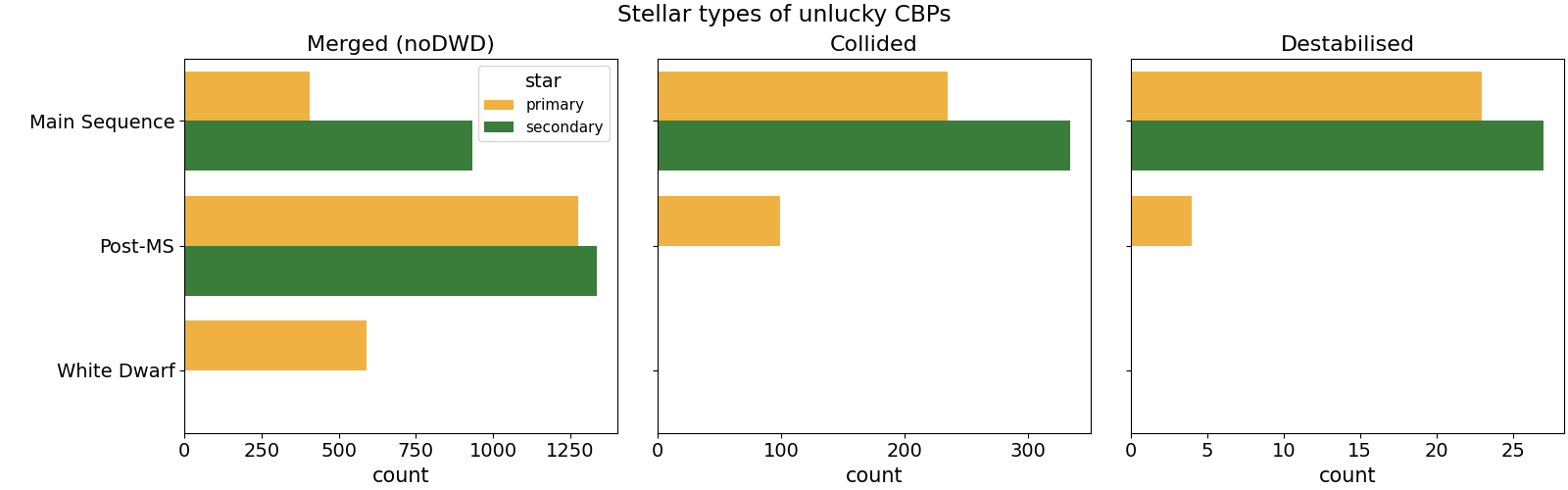}
            \caption{{\it Population A:} Stellar type of the inner binaries, for those planets that do not survive to become Magratheas. From left to right: CBPs around merging binaries, excluded the DWD-mergers, planets in the Collided category and Destabilised. See \secname~\ref{sec:results} for the detailed definitions. The label "Post-MS" includes for simplicity all the stellar types after the Main Sequence and which are not WDs. We show in Fig. \ref{fig:startypes_B} the same plot for Population B.}
            \label{fig:startypes_A}
        \end{figure*}

        It is estimated that the lowest-mass body, or the outermost, of a hierarchical triple, is most often ejected from the system when it becomes dynamically unstable \citep{Busetti+2018, Toonen+2020, Toonen+2022}. In our simulations, the circumbinary gas giants satisfy both conditions of lightest and outermost bodies in the triples, thus they would likely be the ones ejected and transformed into free-floating planets. We provide some number estimates and discussion on the topic in \secname~\ref{sec:discuss}.
        The little statistics of this category did not allow us to produce meaningful plots and histograms and to disentangle the feature of the two populations. However, we note that the majority of the systems become unstable when both stars are in the MS phase (Pop A,  Fig. \ref{fig:startypes_A}, right panel), or when the primary star has evolved in its post-MS phase, before the WD stage (Pop. B, Fig. \ref{fig:startypes_B}, right panel). 
        The common trends we identified in both populations are the followings:
        \begin{itemize}
            \item the planetary semi-major axes are the smallest among all categories (or comparable to the Collided planets) and they shrink from ZAMS to instability; Pop B has a larger semi-major axis lower-bound, the upper bound is the same (\figurename~\ref{fig:cumul_a_out_all});
            \item the inner semi-major axes, on the contrary, are the largest of all the categories, on the opposite of its planetary orbits, setting the conditions for a fragile equilibrium (see \figurename~\ref{fig:cumul_a_in});
            \item the final planetary eccentricity follows loosely the initial distribution, but it has a heavy tail that reaches eccentricities larger than 1, responsible for unbinding the planet. (\figurename~\ref{fig:cumul_e_out});
            \item the stars have several high-mass occurrences, which quickly evolve to disruptive SN. 
        \end{itemize}
    
        Finally, we note that for Pop. A the Destabilised planets all start closer to the critical stability semi-major axis, whereas Pop. B has a wider spread around it. 
        Atmospheric evaporation is completely negligible in this category, likely due to the short lifetime of these triples.

    \subsection{Merged stellar systems}
        \label{sec:res_merged}
    
        \begin{figure}[bt]          
            \centering
            \includegraphics[trim=0.1cm 0.2 0.0cm 0.5 0cm,clip,width=1\linewidth ]{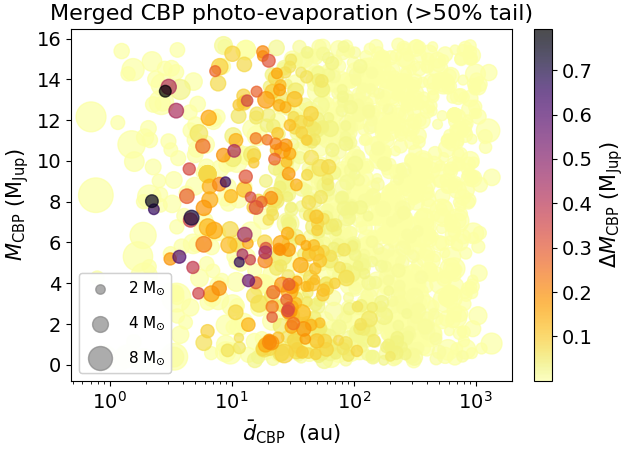}
            \caption{{\it Population A:} atmospheric mass loss endured by gas giants of Merged binaries, scattered in the mass-distance parameter space. The x-axis corresponds to the final time-averaged orbital distance of the CBPs. We show the planets which lost more mass than the 50° percentile of the category. The colour corresponds to the amount of mass lost (as shown in the colour bar). The size of the markers is proportional to the binary progenitors masses, as illustrated in the legend. }
            \label{fig:photoevap_merged_popA}
        \end{figure}
    
        The largest fraction of the simulated systems in our parameter space ends with a merger of the inner binary. The mergers are driven by mass transfer events in the inner binary with the corresponding shrinking of their orbits. Given that mergers are not the focus of this work, we did not further investigate the evolution of the merged star and its planet.
        However, in our isolated environment there is no mechanism to unbind the planet as a consequence of the merger \emph{per se}, for such we only expect an expansion of the planetary orbit as response to the mass loss that takes place during the merger itself.
        Here we present the systems as they were during the last snapshot before the merger moment, to inform on their planetary architecture.
    
        \emph{Population A:} around 32\% of systems had their stars merging within one Hubble Time. 
        \figurename~\ref{fig:cumul_a_out_all} and \figurename~\ref{fig:quad_fused} show how these planets suffered a scatter to large orbital distances, with few of them going beyond $ a_{\rm out} = \SI{1000}{au}.$ 
        However, their semi-major axes are for the most part shorter or comparable to those of Magrathea planets (\figurename~\ref{fig:cumul_a_out_all}).
        The planetary eccentricity distribution is mostly preserved in shape, with some enhancement of the tails, as few planets evolve to cover the whole eccentricity range.
        Photoevaporation has the highest impact for the giant planets in this category (following the Ordinaries), with an average mass loss of 0.2\% and a maximum single loss of 0.9\,\Mj\ (\figurename~\ref{fig:photoevap_merged_popA}). Overall we identified strong evaporation by heavier progenitors than for Magratheas, but the highest mass-loss values are not uniquely related to the shortest final orbital distances.
    
        \emph{Population B:} 35.1\% of the systems merged within a Hubble time. The planet-to-star separation increases similarly to Pop. A, yet their wide orbit abundance decays faster in this population (see third row left panel of \figurename~\ref{fig:quad_fused}). 
        The inner binary eccentricity distribution $e_{\rm out}$ is almost unaltered in time, with mergers happening on the whole range of eccentricities (see third row right panel of \figurename~\ref{fig:quad_fused}).
        Photoevaporation for these planets was weak, with a maximum loss of almost 0.3\,\Mj\ and a mean relative evaporation of 0.02\%.
        
        \emph{Both populations:} 
        We see that the initial distributions of stellar mass in this category cover the whole range of stellar masses, up to the upper limit of our range (10\,\Msun), having larger occurrence rates for the primary stars $M_1$ (see \figurename~\ref{fig:quad_fused}). 
        The binaries semi-major axes are the shortest of all categories (see \figurename~\ref{fig:cumul_a_in}) and their distribution even shrunk in time, for Pop. A, compared to the one of their progenitors (see \figurename~\ref{fig:quad_fused}).
        The Merged category includes hosts at different stellar evolution stages ( \figurename~\ref{fig:startypes_A}), mostly beyond the MS.

        \subsection*{Merged DWD systems}
            \label{sec:dwdmergers}
        
            When further analysing the merged systems categories, we found that a large fraction of Pop. A mergers occur when both stars are white dwarfs. 
        
            \emph{Population A:} 7.5\% of all systems (almost 1/4 of the merged ones) ended up with two white dwarf merging together within one Hubble time. 
            The CBPs semi-major axes in this category spread up to $a_{\rm out} = \SI{1200}{au}$, yet the occurrence of planets decrease at larger planet-to-binary separations. 
            Other features of these objects are similar to the general Merged category, except for the range of stellar mass, which are usually nearly equal mass binaries, and present a lack of the masses above 5\,\Msun,  approximately ( \figurename~\ref{fig:quad_dwdmerged}). 
            The inner binary separation before the merger itself is well below 1\,\Rsun, which is of great relevance for the LISA detectability.
            
            \emph{Population B:} 10.7\% of the simulated systems ends with a DWD merger, more than for Pop. A, reflecting the influence of the giant planets on the final destiny of these triples as a whole. In fact, despite the same distributions have been used to draw stellar binaries in Pop. A and B, the different distribution of the CBPs slightly altered the dynamical equilibrium of the triples and thus the configurations actually initialised and simulated, as a selection effect \citep[see][]{Toonen+2020}. 
            In this population the CBPs semi-major axes reach wide separations too, up to \SI{1500}{au}, with a flatter distribution.

    \subsection{Stable-MT systems}
        \label{sec:res_rlofed}
    
        Among the binary evolution a relevant fraction of binaries (Tab. \ref{tab:percentages}) begun a stable mass transfer process. As previously mentioned the analytic modelling of such a process is not currently included in our software, resulting in a group of systems whose evolution was stopped in an intermediate stage. The difficulty of the simulations lies in the eccentricity of the inner orbit. If such an orbit has not circularised upon the onset of the mass transfer, the mass transfer rate will be orbital-phase dependent, and even episodic, leading to a break down of the classical prescription of mass transfer \citep[but see][]{Sepinsky2007, Sepinsky2009, Dosopoulou2016, Dosoupoulou2016b, Hamers2019}. 
        In circular orbits, stable mass transfer typically leads to the widening of the orbit, when a merger can be avoided \citep[e.g.][]{Soberman1997, Toonen2014}.
         In this regard, these systems may enhance the percentage of Magrathea systems (and/or Destabilised systems) actually present in the Galaxy, at one Hubble Time. 
        
        \textit{Population A:} 16.94\% of the binaries in the total systems begun a stable mass transfer. 
        The CBPs of this category retained an orbit close to the initial one, with a limited scatter to larger orbits (\figurename~\ref{fig:quad_rlofed}). Their eccentricity follows the initial distribution and is weakly dispersed, in line with planets in most categories, as it is illustrated in \figurename~\ref{fig:cumul_e_out}. 
        The average atmospheric evaporation these gas giant suffered is negligible, with a maximum single loss of 0.25\,\Mj.
        
        \textit{Population B:} 17.08\% of the inner binaries in the total systems begun a stable mass transfer. 
        As for Pop. A, the semi-major axes of the circumbinary planets are almost the initial ones, as for the eccentricities (see \figurename~\ref{fig:quad_rlofed}). 
        The atmospheric evaporation from these planets was on average not relevant, with a maximum single loss of 0.13\,\Mj.
    
        \textit{Both populations:} This category present strong similarities in the distributions of both stars and CBPs, reason why we present only one overview panel in \figurename~\ref{fig:quad_rlofed}. 
        The binaries of this group are made of stars of all masses within our range, which, up to the onset of mass transfer, are orbiting at distances equal or smaller than their initial semi-major axes. These stars split up between circular orbits and high-eccentricity orbits, showing a lack of systems in between. 
        As shown in \figurename~\ref{fig:cumul_a_out_all}, the semi-major axes of circumbinary planets in this group are basically always smaller than those of Merged and Magratheas systems planets, and always larger than Collided and Destabilised, lying then in a middle-region between survival and catastrophe.

\section{Discussion}
    \label{sec:discuss}
    
    The modelling of two different populations, A and B, of circumbinary planetary systems show that between 23\% -- 32\% (Tab. \ref{tab:percentages}) of the planets survive the system, becoming Magrathea planets. 
    This result confirm that CBPs can survive the evolution of both host stars, and that such a possibility is not rare. 
    Our results are in agreement with both the discussion by \cite{Kostov16:CE}, and with the results by \cite{FaggingerPortegies22} whose study showed that planets have higher probability to survive one supernova explosion in a binary system, than when orbiting a single supernova progenitor. We infer that planet survival is linked to the stellar mass loss kick, which is lower in the case of compact binary than in the single star case.
    Nevertheless, increasing the initial planet-to-binary separation beyond 200\,au (Sec \ref{sec:CBPpriors}), 
    would reflect in a possible decrease of the fraction of Magrathea planets, due to disruption of the outermost orbits by stellar winds.
    
    In all categories, excluding Collided and Destabilised, we see an expansion of the planetary semi-major axis of a factor $\gtrsim x$4 for both populations A and B, reaching the maximum separation of $a_{\rm out}$ = 1500\,au. 
    The consequence of such an expansion is due to the fact that the CE wind mass loss in \texttt{TRES} is adiabatic, any change of such an assumption could be responsible for a different dynamical response on the outer orbit, hindering the planet survival. 
    While we refer to \secname~\ref{sec:massloss} for a discussion on the topic, we leave the investigation of changes in the wind mass loss and its effect on the Magrathea population for a future work.
    
    Overall, when comparing Pop. A and Pop. B (Fig. \ref{fig:cumul_a_out_all}) we see that Pop. B, i.e., the one initialised  with fully unconstrained planetary priors,  presents larger final separations than Pop. A. Our results also show that there is no selection effect in mass for any of the categories of planets, indicating that the planetary mass does not play a role in the survival of circumbinary gas giants.

    \subsection{Final fate of unstable CBPs}
        \label{subsec:free_floating}
    
        It is important to note that our simulations have not been performed within any Galaxy evolution frame, thus we did not consider the presence of nearby systems, or the effect of external stellar encounters. 
        Such a framework justifies the presence and stability of CBPs at very large separation in our results. 
        Any change in these assumptions (together with non-adiabatic winds, see \secname \ref{sec:massloss}) could be responsible for either the capture of weakly gravitationally bound planets by external systems, or planetary loss to Galactic tides.
         
        Dynamical instabilities in planetary systems can be a source of \emph{free-floating} planets (or also {\it rogue} planets), unbinding the SSO from the host star(s) and leaving it wandering alone in the interstellar space. 
        This type of planets was predicted by planet formation models \citep{VerasRaymond2012,Ma2016} and they were first discovered in star-forming regions \citep{ZapateroOsorio2000,Luhman2005,Marsh2010}. 
        Their abundance in the Milky Way, and typical mass scale remains uncertain. Consequently, their origin is doubtful, and their formation mechanisms remain an open question. 
        Binary stars, however, could be an efficient source of free-floating planets, as it emerged from the theoretical studies by
        \cite{Sutherland&Fabrycky2016:unstables} and \cite{Smullen+2016}. These authors simulated the evolution of unstable circumbinary planets, with N-body calculations, covering different parts of the parameter space. 
        They found that in the majority of cases an unstable CBP will be ejected from the system, much more often than planets around single stars. 
        In particular, \citet{Sutherland&Fabrycky2016:unstables} found ejection rates ranging from $\sim$80\% to 93.5\% and collision rates on the secondary star from 3.6\% to $\sim 8\%$, depending on the binary properties. The collisions happen less often on the primary star. 
        The consecutive study by \citet{Smullen+2016} reported results consistent with \citeauthor{Sutherland&Fabrycky2016:unstables}. 
        Among our results, we have unstable CBPs in the Collided and Destabilised categories. To provide a rough estimate, assuming that 80\% of the unstable planets are indeed ejected, our simulations would generate $\sim 240 $ free-floating planets over 10500 total (averaged between Pop. A and B). 
    
        The collisions are significantly more rare, as reported above, but they deserve a special interest of their own.
        In general, if a collision occurs in the last stages of the stellar life, when the star is a WD, it should pollute its atmosphere with planetary material. This effect is widely observed nowadays (e.g., \citealt{Zuckerman2010,Koester14:WD,Wilson2019:pollut,Veras2021} and references there-in).
        In our Collided category (Pop. B, \figurename~\ref{fig:startypes_B}) we have only three systems where the primary is a WD and for such we cannot extrapolate any meaningful statistics on the occurrence of similar events. With a larger simulation sample we could theoretically start constraining it.

    \subsection{Systems which hit CPU limit}
        \label{sec:cpu_limited}

        Having set a maximum computational time for each step in our simulations, a portion of our systems did not run to completion. A total of 12\% and 2.5\% of systems hits the CPU limit in Pop. A and Pop. B, respectively. 
        These systems appear to be lying in regions of the parameter space unsuited for the secular approximation, which tries to reduce the integration time to smaller steps until breaking the CPU time limit. Many of these systems are piled up onto the stability limit (\figurename~\ref{fig:semiaxes_ratios}, bottom panels), so that any deviation could lead to wild dynamical evolution and likely phases of instability. 
        The inner binaries show rather high eccentricities and small orbital separations overall, reinforcing the picture of inadequacy to be simulated in the secular regime. 
        Given that their computation stopped, we cannot predict the final parameter space outcome for these systems. 
        However, judging on their proximity to the instability region, it is reasonable to assume that these systems would hardly survive until one Hubble Time to become Magratheas, instead they would probably end up within the Destabilised or Collided categories, the fate of which remains ultimately uncertain. 
        Based on these remarks, we can affirm that the CPU-limited systems have a negligible impact in characterising the features of the surviving CBPs population, which was the ultimate goal of this work.
        As a matter of fact, we see that CPU-limited systems share a similar distribution of period ratios with Collided and Destabilised: for all these three categories, the great majority of planets orbit with periods ratios close - or within 10 times - the critical ratio for stability,  P$_\mathrm{crit} $, (defined as the period-equivalent of the critical semi-major axis given by \eqname~\ref{eq:stability}); P$_\mathrm{crit} $ itself peaks at values from 6 to 8. For comparison, Magrathea planets have periods ratios which are a factor 10 -- $ 10^6 $ times their P$_\mathrm{crit} $, which instead peaks before the P$_\mathrm{crit}$ = 4 value. 
        From \figurename~\ref{fig:semiaxes_ratios} we can see the same argument from the critical semi-major axis perspective (which is solely a function of the inner binary parameters, Eq. \ref{eq:stability}): while Magrathea planets show a large vertical spread for the semi-major axes ratio (\nicefrac{a$_{\rm out~}$}{~a$_{\rm in}$}, in unit of a$_{\rm crit}$), in correspondence of a$_{\rm crit} \approx 2.39 $ (top panels), both Collided and CPU-limited categories have systems that accumulate at larger values of a$_{\rm crit} \approx 4 $ and present a smaller vertical spread (second row and last row panels, respectively).
        Moreover, CPU-limited systems mainly consist of Main Sequence stars and they do not include any DWD binary.
        With regard to this aspect, under the observational point of view we see a pile-up close to the stability boundary of circumbinary planets orbiting Main Sequence stars \citep{Martin2018,Martin20}, similar to the pile-up and stellar evolutionary types that what we see within the CPU-limited category. Such result suggest that the known close-in CBPs, detected by transit, will most likely not survive the evolution of the binaries and will end their life by colliding or being ejected from the system. The fact that most of the CBP population known today will not survive, it further justify the need for dedicated surveys to hunt for long period circumbinary exoplanets with RV (e.g., BEBOP, \citealt{Martin2019:bebop}), or direct imaging.
    
        All considered, given that the CPU-limited systems evolution is computationally expensive, we save for future studies their analysis without a CPU time limit, to  characterise their nature, as we did here with the rest of our sample.

    \subsection{The effects of mass loss}
        \label{sec:massloss}
        
        In this study we have assumed that any mass loss from the system affects the orbits in an adiabatic way. Our results, e.g.,the 23-32\% fraction of surviving planets,  are therefore strictly only valid under this assumption. How good is this assumption? 
        
        An example of a process responsible for mass loss in a stellar system are the stellar winds. For low- \& intermediate mass stars the wind mass loss predominantly occurs late in the evolution of the star during the giant phases. The assumption that the wind matter is lost from the system in an adiabatic way is a common assumption to make in binary evolutionary calculations \citep[see e.g.,][for a comparison of four binary evolutionary codes]{Toonen2014}, however, that does not make the assumption justified. \cite{Toonen2017} demonstrate that the assumption breaks down in binary orbits larger then a few thousand au. In 2011, the seminal work of \cite{Veras11} already showed the break down of the adiabatic assumption in the planetary context, that is for wide-orbit circumstellar planets or Oort-cloud analogues. They find that up to 20\% of Oort-cloud planets at $10^5$au are ejected due to the non-adiabaticity of the stellar winds, and planets at $\approx 10^4$\,au may experience a moderate eccentricity change of a few tenths. 
        
        The reason for the breakdown is that the timescale of the mass loss becomes comparable or significantly shorter than the orbital period. In this case the mass loss rate is no longer constant during a single orbital revolution, but becomes phase-dependent. Whereas adiabatic mass loss merely leads to a widening of the orbit \citep[see e.g.][]{Rahoma09}, phase-dependent mass loss may lead to the dissolution of the orbit. This can be easily understood by considering the limit of instantaneous wind mass loss as an analogue of supernova explosions  which are well known for their destructive effect on binary orbits \citep[see e.g.][]{Hills83}.
        
        To estimate the effect of non-adiabatic mass loss on our results, we have to consider the initial orbital configuration of our planetary objects. In the current study, we only consider planetary orbital separations of less than 200\,au, see Section \ref{sec:CBPpriors}. Therefore the adiabatic assumption in our modelling of the stellar winds seems justified, and we do not expect that our planetary orbits will be dissolved (or the eccentricities to be affected) significantly. 
        
        Another process that can lead to mass loss in a binary system is non-conservative mass transfer. Within the context of our simulations set-up, this would be the mass loss during a common-envelope event. As discussed in section \ref{sec:CE}, the physics behind the CE-mass ejection is actively debated. As hydrodynamical simulations typically do not unbind the envelope,  \citep[ with only a handful of  exceptions e.g.][]{Ivanova2016, Law2020}, the computational constraints on the velocity of the escaping material (or the timescale of ejection) is limited. One may expect ejection on a dynamical timescale, however, observational indications hint at much longer timescales,  of the order of $\sim 10^4$ yr \citep{Michaely19, Igoshev20}. If we assume that a given  orbit with orbital period larger than $10^5\, (10^4)$ yr is disrupted during a CE-event (i.e., the mass loss occurs during less then 10\% of a single revolution), it would mean that circumbinary planets with orbital separations larger than 3000-4000\,au (600-1000\,au) would be lost from the system. Given our maximum initial orbital separation of 200\,au for the planets, our simulated systems are not typically disrupted by CE mass loss even in the case of non-adiabatic mass loss.

    \subsection{The LISA framework}
    
        \begin{table}[tb]         
            \centering
            \begin{tabular}{l  r r  r r }
                \toprule
                 & \multicolumn{2}{c}{$ P < 10 $ yr  }   & \multicolumn{2}{c}{$ P < 50 $ yr  }  \\
                \cmidrule(l{.75 em}r{.75 em}){2-3}
                \cmidrule(l{.75 em}r{.75em}){4-5}
                 &  \multicolumn{1}{c}{Pop. A}  &  \multicolumn{1}{c}{Pop. B}    & \multicolumn{1}{c}{Pop. A}  &  \multicolumn{1}{c}{Pop. B} \\
                \midrule
                Magrathea  &   0.00 \%    &   0.00 \%    &   0.08 \%    &   0.03 \% \\
                Collided    &   69.46 \%     &   9.91 \%    &   92.51 \%    &   25.68 \%  \\
                Destabilised    &   59.26 \%    &   11.11 \%  &   77.78 \%    &   16.67 \%   \\
                Merged      &   8.05 \%    &   0.33 \%   &   20.55 \%    &   2.88 \%  \\
                Stable-MT   &   26.42 \%    &   1.95 \%   &   46.32 \%    &   7.81 \%  \\
                CPU-limited   &   72.40 \%    &   19.31 \%   &   92.15 \%    &   45.56 \%  \\
                Ordinaries   &   12.73 \%    &   1.07 \%   &   31.26 \%    &   4.80 \%  \\
                \bottomrule
                \end{tabular}
            \caption{Percentages of CBPs with periods within 10\,yr and 50\,yr, divided by category and population. 
            All categories show significantly higher percentages for Population A than B, reflecting the different initialisation.}
            \label{tab:shortPeriods} 
        \end{table}

    The gravitational wave frequency $f_{\rm GW}$ of a stable monochromatic circular binary, consequently valid for DWD, can be approximated to the first order as  $f_{GW}$ = $\nicefrac{2}{P_{\rm bin}}$ \citep{Korol2020}. 
    Note that the timescale on which the orbital period of a typical DWD changes, due to the GW emission, is significantly longer than the LISA nominal mission lifetime. The approximation of $f_{GW}$ excludes the back reaction produced by the GWs and it hence considered valid for Galactic DWD.
    The amplitude $\mathcal{A}$ of the signal is given by
    
    \begin{equation}\label{eq:amplitude}
        \mathcal{A} = \frac{2 (G \mathcal{M})^{\nicefrac{5}{3}}(\pi f_{GW} )^{\nicefrac{2}{3}}}{c^{4}\ d}
    \end{equation}
    where $G$ is the gravitational constant, $c$ is the speed of light, $\mathcal{M}$ is the chirp mass of the inner binary $\mathcal{M}= (M_1 \cdot M_2)^{3/5}/(M_1 + M_2)^{1/5}$ and $d$ is the distance of the binary system from the observer. 
    The characteristic strain of these binaries is defined as $h_c = \mathcal{A} \sqrt{f_{GW}\cdot T_{\rm obs}}$ \citep{Korol2020},
    where $T_{\rm obs}$ is the LISA mission lifetime of 4 years.

    The DWD mergers happening within a Hubble Time (\secname~\ref{sec:dwdmergers}) are particularly interesting as they will pass through the LISA frequency band. This means that the GW that these compact DWDs usually produce, will change its frequency as function of shrinking of the orbit towards the merging event;  when their period falls between 3 to 60 min they can be detected by LISA \citep{Korol2017}. In our Merged-DWDs category the binary period distribution peaks around $ 10 $\,min (in semi-major axis: $a_{\rm in} \sim 5\cdot10^{-4}$\,au $\sim 0.1$ \Rsun, Fig. \ref{fig:binsep_magra_dwdmerged}), falling well within the LISA band. 
    Each of the merging DWD hosts a planet before merging.
    In the moment the binary is emitting in the LISA band (i.e., when P$_{\rm bin} < 60$ min) such a planet has the potential to be detected by LISA, depending on the observing time, 
    the planetary mass, semi-major axis, and highly on the signal-to-noise of the GW \citep{TamaniniDanielski19:CBP,Danielski19:CBP}.
    Moreover, LISA has the potential to detect those giant planets (M$_P >$ 0.2 \Mj) whose period is shorter than the LISA observing period T$_{\rm obs} \approx 4$\,yr, considering a formal duty cycle of 100\% and ignoring maintenance operations and data gaps. 
    A recent Bayesian analysis by \cite{Katz2022}, who studied in detail the possibility of detections of planets with eccentric orbits, showed that the posteriors behaviour of these triple systems is highly dependent on both the detectability of the GW source and the period of the third-body orbit. More specifically, CBPs periods shorter than $P = T_{\rm obs}/2$ present planetary posteriors that remain fairly Gaussian, increasing the ability to resolve the period of the planetary perturber. This does not prevent some of longer periods planets ($\nicefrac{T_{\rm obs}}{2} < P < T_{\rm obs}$) to be detected but, given the lack of Gaussianity of their posterior, due to the failure to meet the Nyquist criterion on the planetary orbit sampling, it is harder to constrain their period.

    To this regards, looking specifically at the category of merging DWDs, we have no systems in both populations with planetary periods shorter than 4 years. 
    More specifically, the shortest planetary period is \SI{13.8}{yr} (4.6\,au), and \SI{96.6}{yr} (\SI{17.2}{au}) in Pop. A and B, respectively.
    Similarly, for the Magrathea systems the shortest planetary period is \SI{43.4}{yr} (9.6\,au) in Pop. A, and \SI{20}{yr} (6\,au) in Pop. B. We remark that Pop. B has almost one thousand more Magratheas than Pop. A, highlighting that the lack of planets with LISA-compatible periods might result from statistical biases due to the limited number of systems simulated. The smooth decay of the periods distribution together with the evidence that, with an higher number of Magrathea planets we observed shorter periods, suggest that single giant planets on close-orbits are indeed a possibility, even if rare. \\
    We report in \tablename~\ref{tab:shortPeriods} the percentages of CBPs with orbital period smaller than 10 (corresponding to the length of a possible extended LISA mission) and 50 years (as additional reference), for all the categories in both populations.
    It is important to stress that the numbers in \tablename~\ref{tab:shortPeriods} are the result of the evolution of a single planet circumbinary population, and do not represent what it could be the dynamics of a multi-planet system. In the latter case, destabilisation and dynamical chaos, for instance the dynamic of planet-planet scattering \citep{RasioFord,WeidMarzari96}, might occur as a consequence of binary evolution, and can be responsible of changing the orbit of the planets in the system, migrating them inwards, or pushing them outwards. 
    Furthermore, in our simulations we are not including interactions with the circumbinary disc that form as a consequence of the common envelope phase \citep{KashiSoker2011,Passy2012}. The presence of a  second-generation disc can be in fact responsible for inward migration of the surviving planet \citep{Perets2010} similarly to what is theorised during planetary formation  \citep[e.g.,][]{Johansen2019,Tanaka20}. The external radius observed for circumbinary discs around evolved binary is usually found between a few dozen au to 500\,au \citep{Rafikov2016, IzzarJermyn}, which covers mostly all range of separations of our CBPs.
    A third possibility, for a planet to reach inner orbits, and already described by \cite{Perets2010}, is the interaction among first- and new generation planets. What it was a previously steady planetary configuration may dynamically evolve into a new configuration as a consequence of the formation of new planets in a post-CE disc \citep{SchleicherDreizler2014,Ledda2023}, leading again to several outcomes like inward migration and/or ejections and/or collisions with the binary. 
    Dedicated studies will have to be performed to explore the various outcomes of such evolutionary paths, to better constrain the properties of the {\it surviving} first generation planets that LISA will be able to detect, 
    in the light of the formation of {\it second-generation} planets beyond the MS, for instance around DWD binaries, where different type of planets can form and migrate within 1\,au from the inner binary \citep{Ledda2023}.

\section{Conclusions}
    \label{sec:conclude}

    We studied with numerical simulations the long-term evolution of circumbinary planetary systems hosting a single giant planet. To do so we generated two populations of 10500 systems with different planetary priors. The first population (A) had the planets initialised based on the distributions of known planets in single star systems, as reported in the literature; the second population (B), had the planets initialised with uniform distributions, to serve as unbiased comparison benchmark. The priors for the inner binaries were chosen based on the models yielding the best agreement with the observations of DWDs in the Galaxy \citep{Toonen12:DWD, Toonen2017}, with stellar masses ranging from 0.95 to 10\,\Msun. We employed the assumption of adiabatic stellar winds and the model of \cite{Nelemans01:DWD} for the CE phase.
    We eventually divided the simulated systems based on their evolutive outcome, with the ultimate objective of characterising the \emph{Magrathea} planets: the planetary survivors orbiting DWDs within one Hubble time. We found a relatively high occurrence of Magratheas in our sample, which gives optimistic hopes for future observations targeted on this category. All our systems have been simulated in an isolated environment, outside the framework of any Galactic environment. 
    
    The data obtained from our simulations can be analysed from different perspectives and can answer a variety of scientific questions. We focused here on the general evolution of these systems in the orbital parameter space, with particular attention to the CBPs closest to their stars. 
    We summarise below the main takeaways of our data analysis:
    
    \begin{itemize}
        \item 23\% to 32\% of all giants survive for one Hubble Time to become Magrathea planets (respectively Pop. A and B.) and they tend to reside on wide orbits around light hosts, regardless of their eccentricity;
        \item the largest fraction ($\sim 33\%$) of the systems ends up with a merger in the inner binary, which involves DWD mergers for more than one quarter of them;
        \item the CBP mass, in the simulated giant planet range, does not appear to be correlated with its host system category, and does not play a role in the survival;
        \item photoevaporation has a negligible impact on the majority of our giant planets, due to the adiabatic expansion of their orbits during every stellar wind phase;
        \item CBPs prefer prograde orbits during their evolution, especially the Magratheas;
        \item unstable systems are short-lived, they can potentially create a few percents group of free-floating planets and set the conditions for WD pollution;
    \end{itemize}

    Within the framework of development of the LISA science case for the detection of exoplanets we found that the evolution of single giant planet systems do not produce any systems in the range of sensitivity of LISA. In fact, no planet is found with a period smaller than 20\,yr among the Magratheas, and than 14\,yr in the Merged-DWDs. Our results are a direct reflection of the number of systems we have simulated, 
    and much larger simulations need to be performed to check whether they can relieve the statistical bias. 
    
    Finally, the variables that play a role into the evolutionary path of a planetary system are many. The study here presented aims at being the first step towards a more comprehensive analysis 
    which accounts for additional physical processes occurring at different stages of the binary life. Dedicated set of evolutionary studies, that include the presence of multiple planets, the presence of a post-MS disc, and the possible interaction between first- and new generations planets \citep{Ledda2023}, needs to be further developed to better address the final fate of circumbinary systems, and hence the occurrence rates of the LISA exoplanets.

\begin{acknowledgements}
    The authors thank Antonio Claret for providing us with the apsidal motion constant models. We also thank Eva Villaver, Valeryia Korol, Diego Turrini and Dimitri Veras for the helpful discussion. 
    G.C. and C.D. acknowledge support from the Ministerio de Ciencia e Innovaci\`on, through the project ACERO Ref. PID2019-110689RB-I00/AEI/10.13039/501100011033;
    C.D. acknowledges financial support from the grant CEX2021-001131-S funded by MCIN/AEI/10.13039/501100011033.
    S.T. acknowledges support from the Netherlands Research Council NWO (VENI 639.041.645 and VIDI 203.061 grants).
    Some of the computations described in this paper were performed using the University of Birmingham's BlueBEAR HPC service, which provides a High Performance Computing service to the University's research community. See \url{http://www.birmingham.ac.uk/bear for more details}.    
\end{acknowledgements}

\bibliographystyle{aa}
\bibliography{CB_References}

\begin{appendix}  

\section{Additional plots}

    We include in the Appendix additional plots representative of our results. The first block of plots consists of the overview of the distributions for all the categories described in \secname \ref{sec:results}.
    Then, we present plots relative to the Population B: the cumulative CBPs eccentricity plot,  the inclination distribution comparison plot, and the stellar types of unlucky CBPs both for Population B.
    Lastly, we report a series of plot concerning the critical semi-major axis scatter of CBPs in the various categories.


    \begin{figure*}[htb]          
        \centering
        \includegraphics[width=\linewidth ]{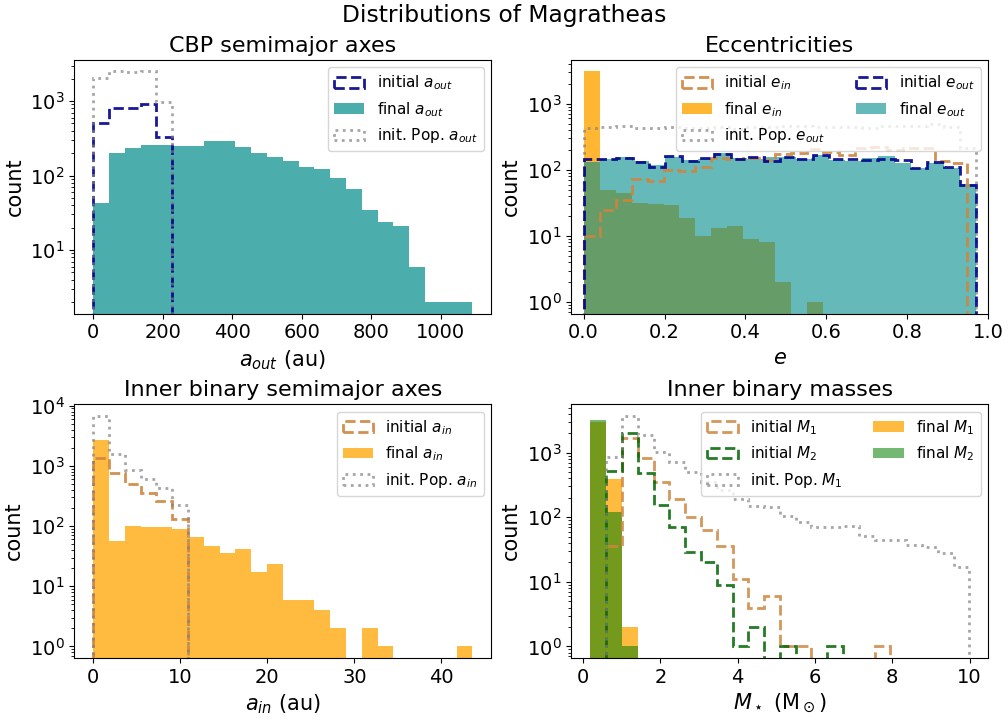} 
        \caption{{\it Population B:} overview of the distributions of Magrathea systems in the parameter space. Solid histograms represent the Magrathea parameters at one Hubble time, while the dashed lines show their initial distributions. The "out" subscript denotes the planetary parameters (blue distributions). The dotted grey lines, instead, show the initial distributions for the whole population, not restricted to the Magratheas.}
        \label{fig:quad_magratheas_B}
    \end{figure*}

    \begin{figure*}[t]          
        \centering          
        \includegraphics[width=0.85\linewidth ]{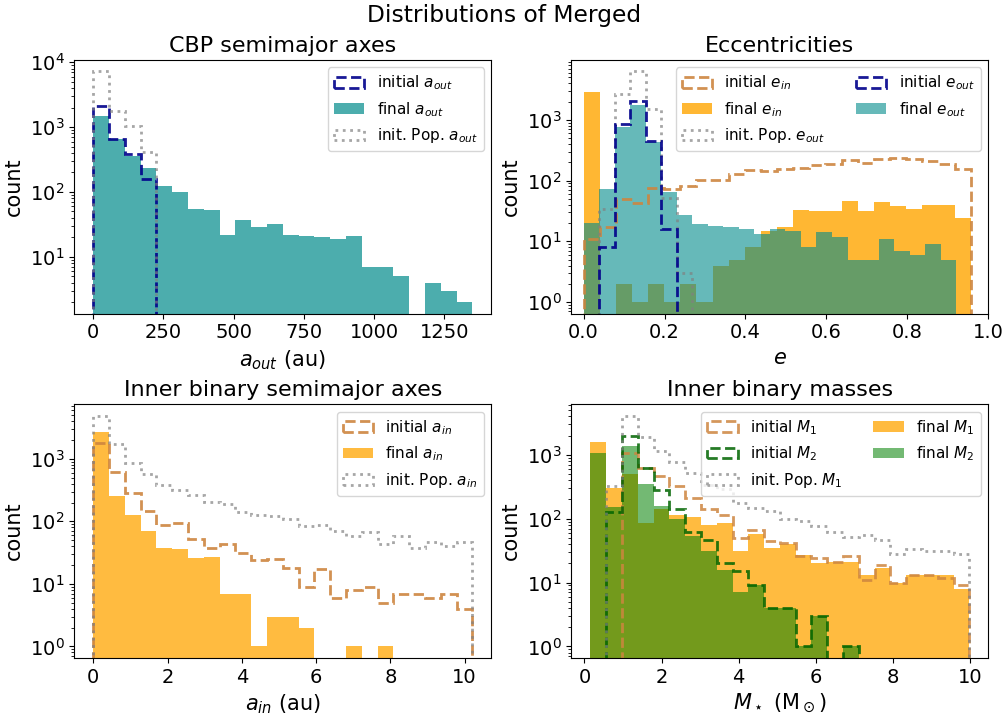} \\
    
        \bigskip
        
        \includegraphics[width=0.85\linewidth ]{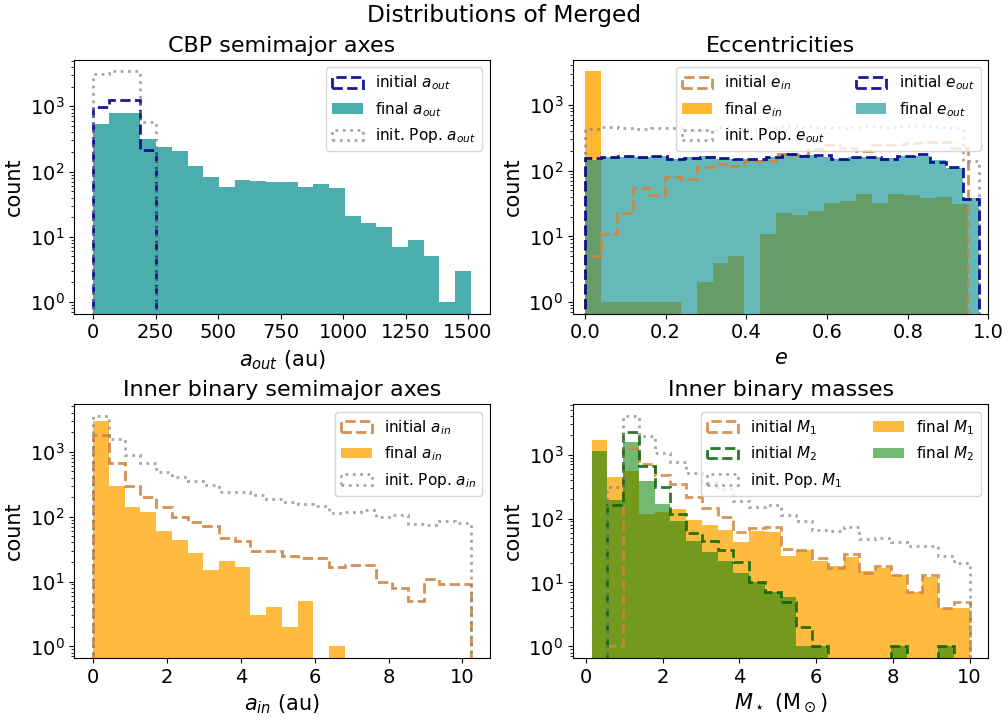}  
        \caption{ overview of the Merged systems distributions. Solid histograms represent the Merged parameters at one Hubble time, while the dashed lines show their initial distributions. The "out" subscript denotes the planetary parameters (blue distributions). The dotted grey lines, instead, show the initial distributions for the whole population, not restricted to the Merged. \emph{Top half}: Pop. A, \emph{bottom half}: Pop. B.
        Note that for these systems the parameters are those referring to the last moment preceding the inner binary merger. }
        \label{fig:quad_fused}
    \end{figure*}

    \begin{figure*}[t]          
        \centering          
        \includegraphics[width=0.85\linewidth ]{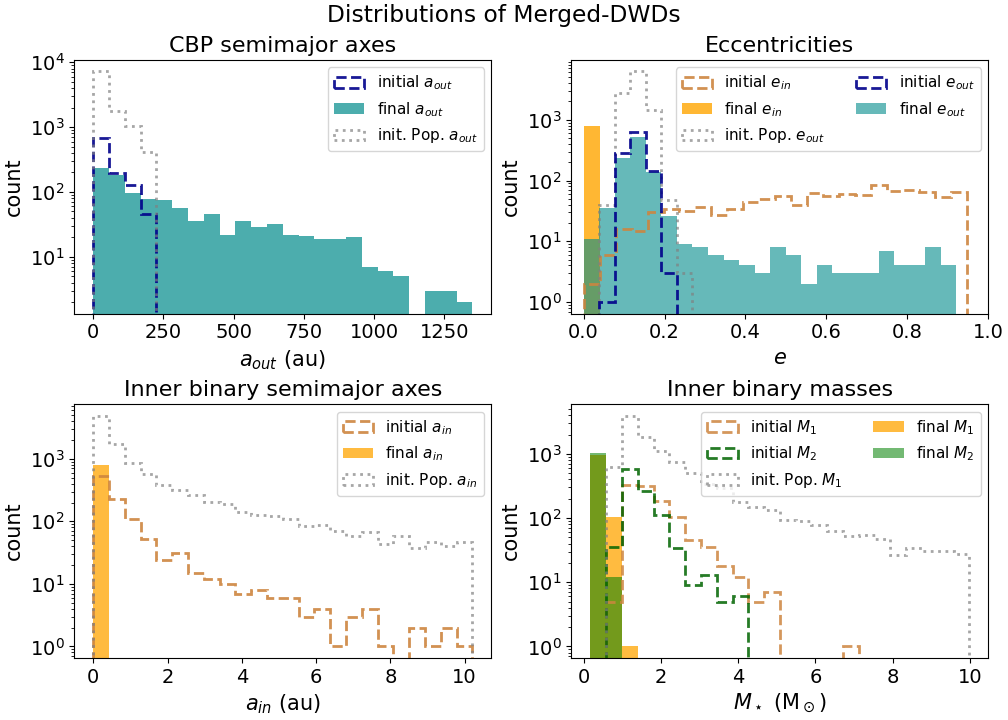} \\
    
        \bigskip
        
        \includegraphics[width=0.85\linewidth ]{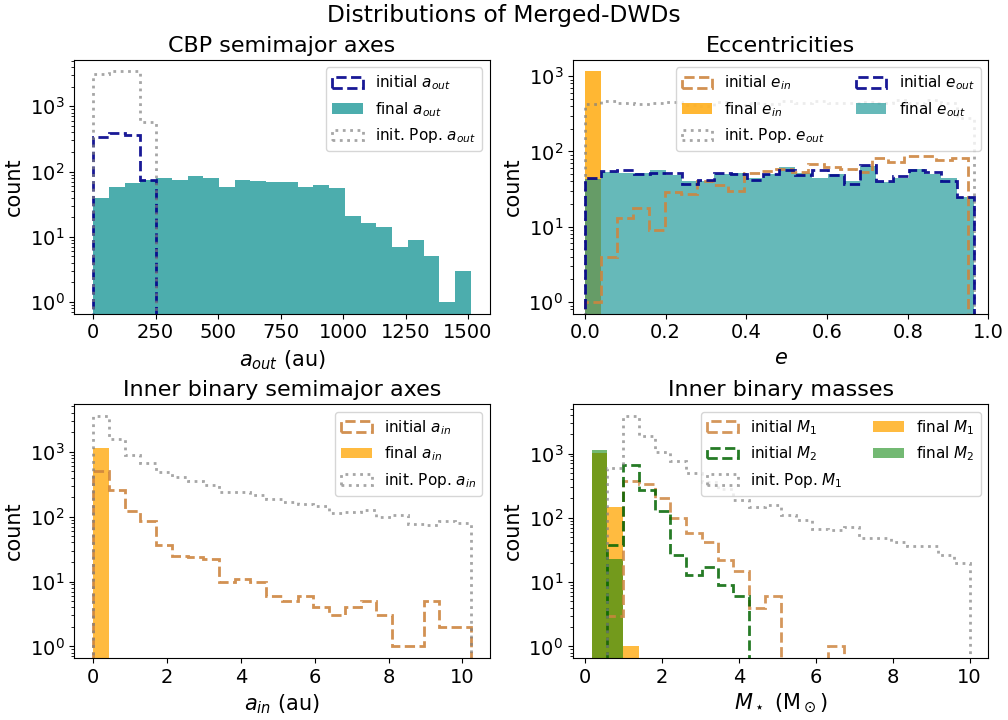}  
        \caption{ overview of the Merged-DWDs systems distributions. Solid histograms represent the Merged-DWDs parameters at one Hubble time, while the dashed lines show their initial distributions. The "out" subscript denotes the planetary parameters (blue distributions). The dotted grey lines, instead, show the initial distributions for the whole population, not restricted to the Merged-DWDs. \emph{Top half}: Pop. A, \emph{bottom half}: Pop. B.
        Note that for these systems the parameters are those referring to the last moment preceding the inner binary merger. }
        \label{fig:quad_dwdmerged}
    \end{figure*}

    \begin{figure*}[t]   
        \centering
        \includegraphics[width=0.85\linewidth]{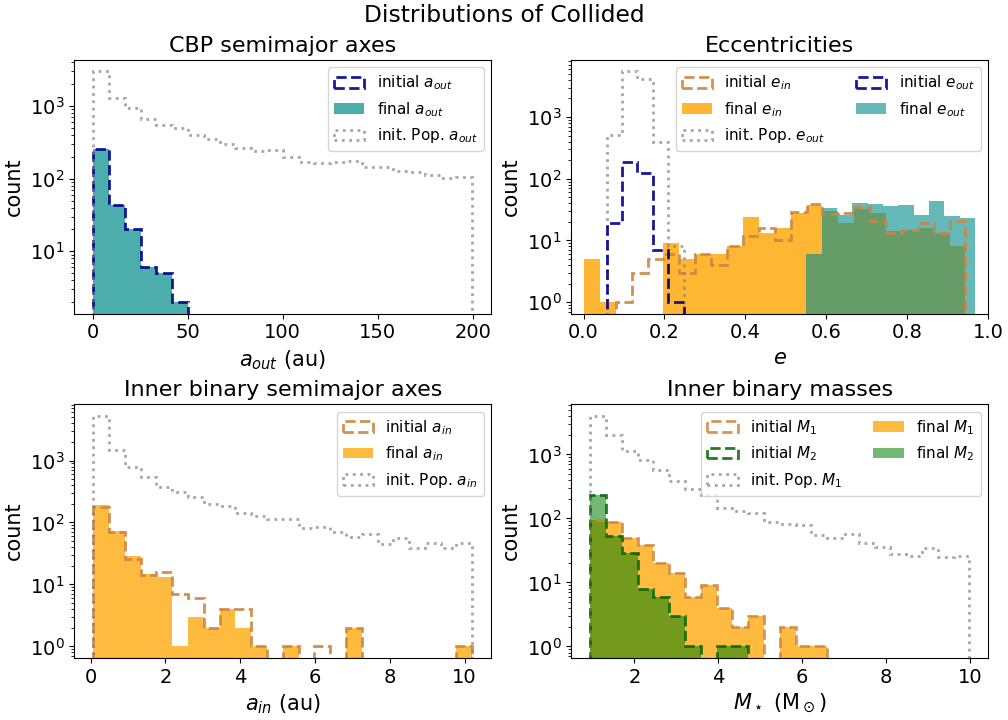} \\
    
        \bigskip
        
        \includegraphics[width=0.85\linewidth]{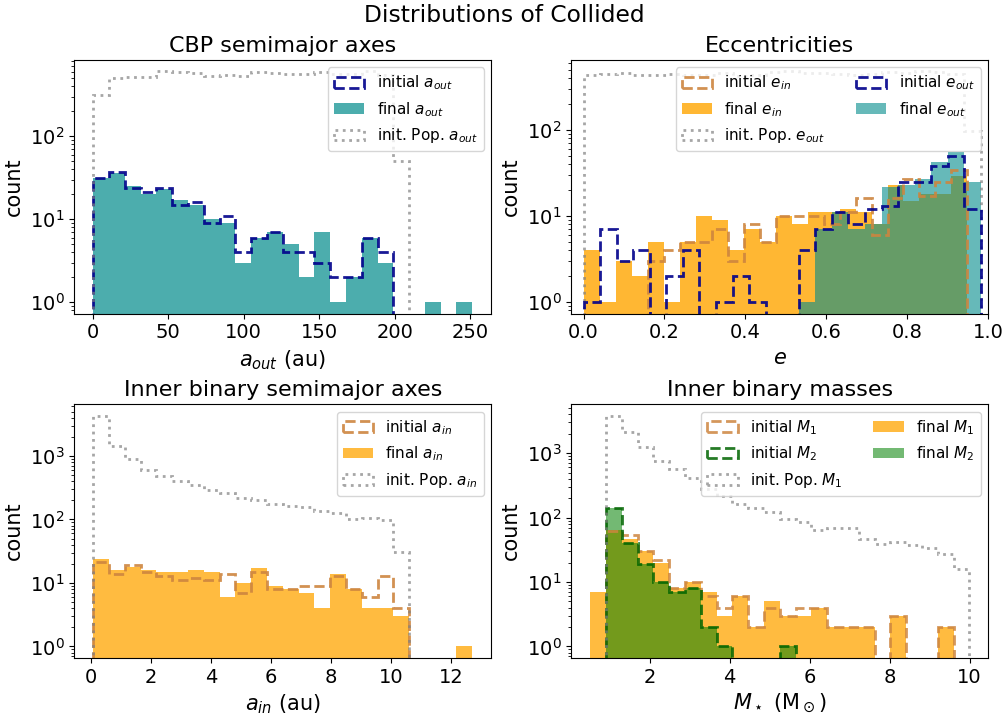}
        \caption{ overview of the Collided systems distributions. Solid histograms represent the Collided parameters at one Hubble time, while the dashed lines show their initial distributions. The "out" subscript denotes the planetary parameters (blue distributions). The dotted grey lines, instead, show the initial distributions for the whole population, not restricted to the Collided. \emph{Top half}: Pop. A, \emph{bottom half}: Pop. B.}
        \label{fig:quad_grilled}
    \end{figure*}

    \begin{figure*}[t]          
        \centering          
        \includegraphics[width=0.85\linewidth ]{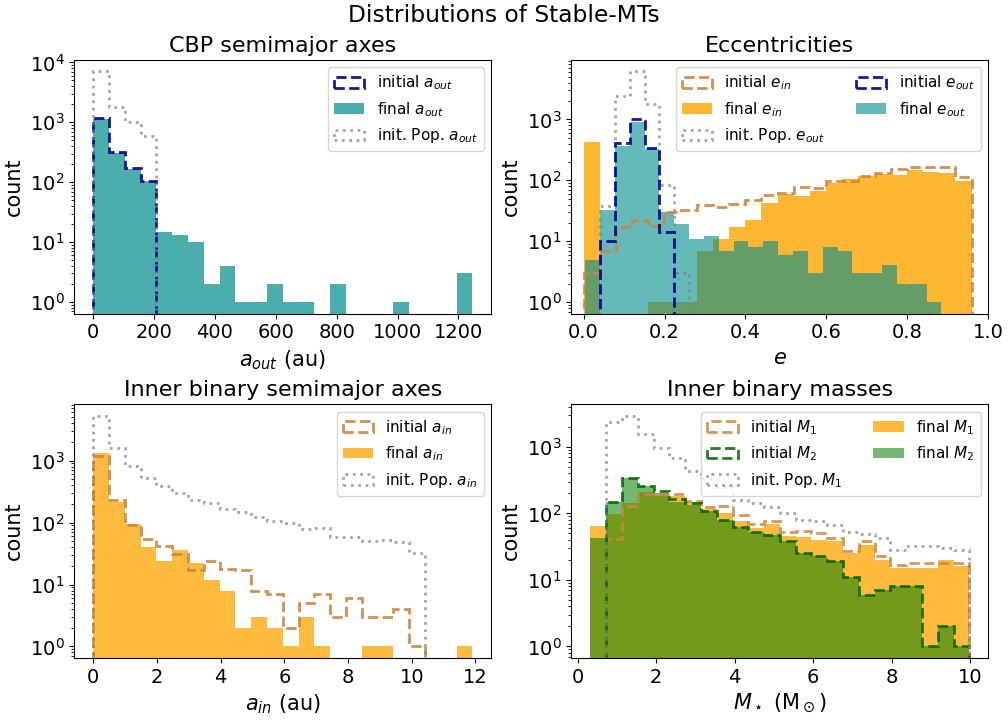} \\
    
        \bigskip
            
        \includegraphics[width=0.85\linewidth ]{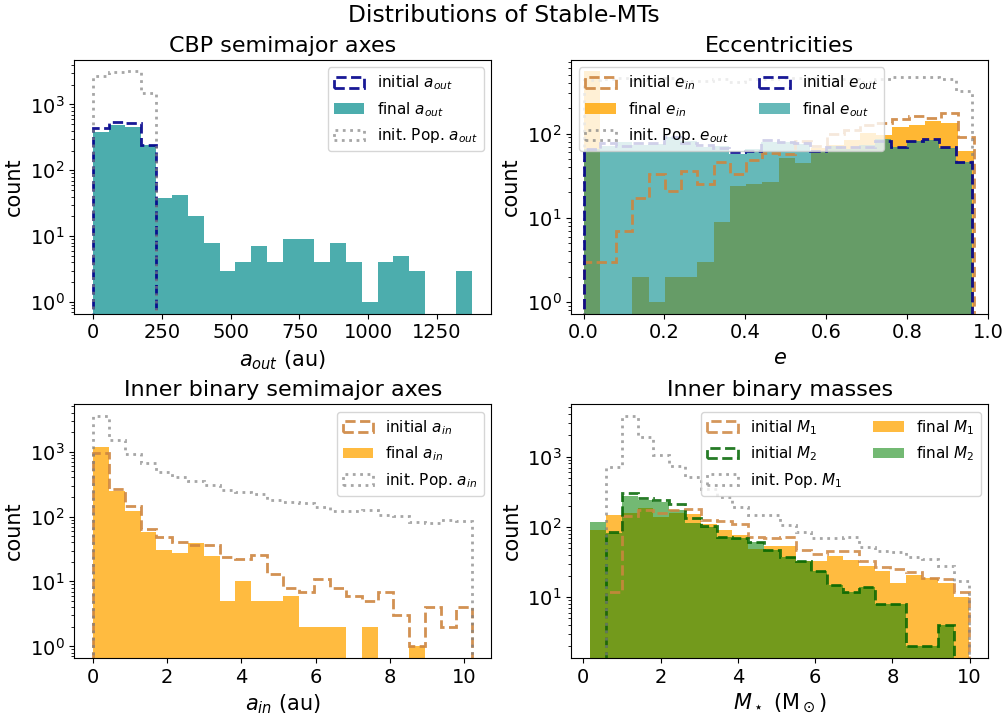}  
        \caption{ overview of the Stable-MT systems distributions. Solid histograms represent the Stable-MT parameters at one Hubble Time, while the dashed lines show their initial distributions. The "out" subscript denotes the planetary parameters (blue distributions). The dotted grey lines, instead, show the initial distributions for the whole population, not restricted to the Stable-MT. \emph{Top half}: Pop. A, \emph{bottom half}: Pop. B. }
        \label{fig:quad_rlofed}
    \end{figure*}

    
    \begin{figure*}[t]          
        \centering
        
        \includegraphics[width=0.5\linewidth ]{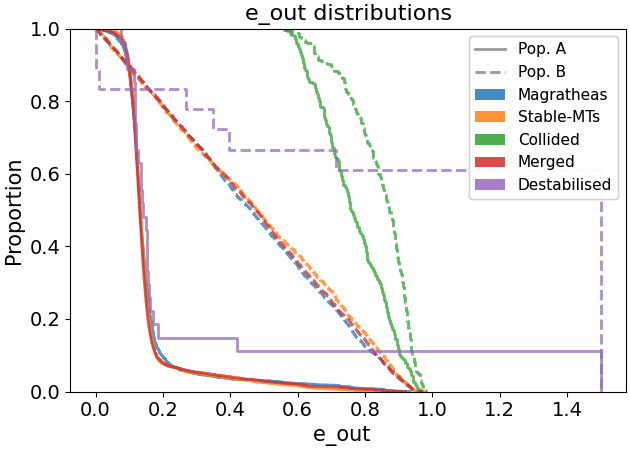} 
        \caption{Cumulative distributions for eccentricity of planets. \emph{Solid line}: Pop. A, \emph{dashed line}: Pop. B. \\ 
        The eccentricity of those Destabilised systems with unbound orbits ($ e > 1 $) has been artificially set to 1.5 to avoid useless dispersion in the plots.
        The Collided systems stand out from all the other categories, for both populations.}
        \label{fig:cumul_e_out}
    \end{figure*}

    
    \begin{figure*}[t]          
        \centering
    
        \includegraphics[width=0.5\linewidth ]{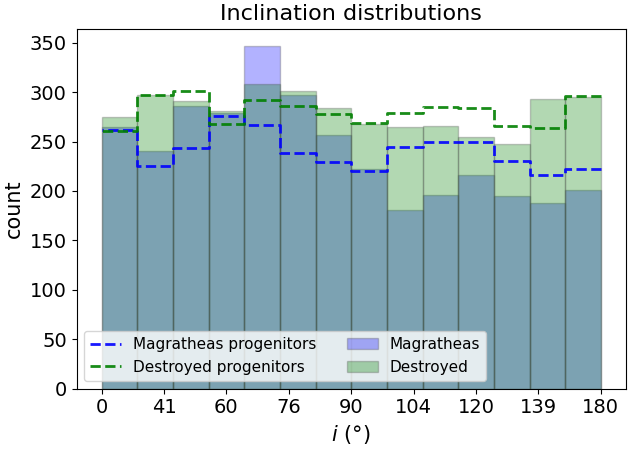} 
        \caption{{\it Population B:} Relative inclination distributions for Magrathea and the destroyed systems (Merged, Collided, Destabilised). }
        \label{fig:PopB_hist_inclination}
    \end{figure*}

     \begin{figure*}[t]       
            \centering
            \includegraphics[width=\linewidth ]{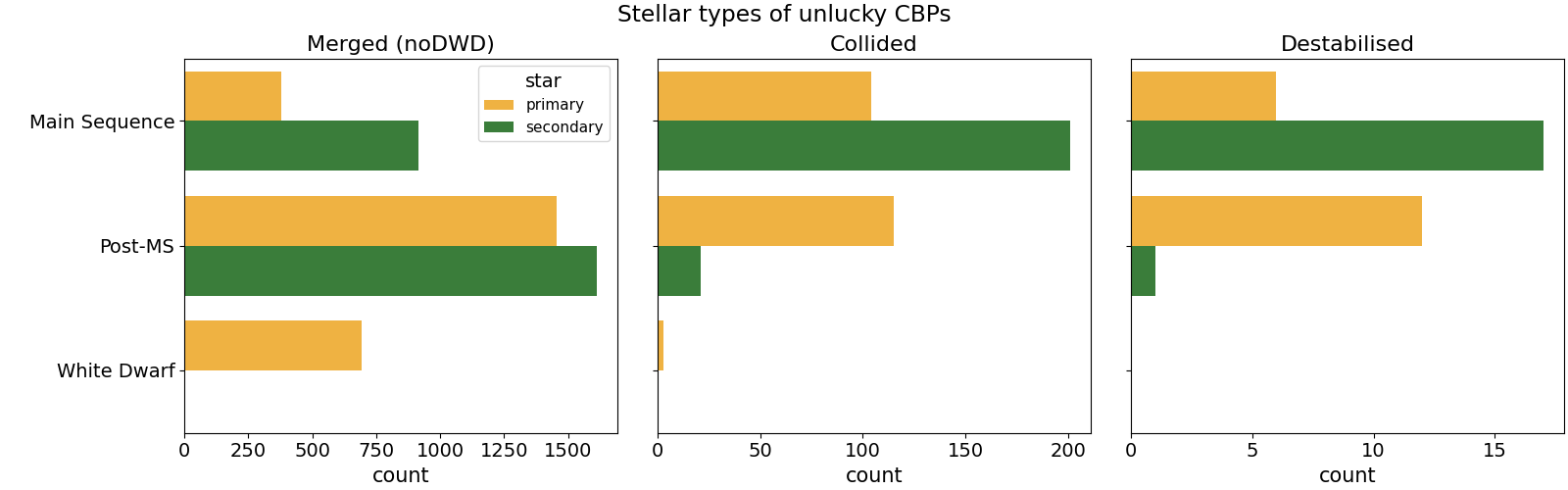}
            \caption{{\it Population B:} Stellar type of the inner binaries, for those planets that do not survive to become Magratheas. From left to right: CBPs around merging binaries, excluded the DWD-mergers, planets in the Collided category and Destabilised. See \secname~\ref{sec:results} for the detailed definitions. The label "Post-MS" includes for simplicity all the stellar types after the Main Sequence and which are not WDs. }
            \label{fig:startypes_B}
        \end{figure*}

    
    \begin{figure*}[t]          
        \centering
        \includegraphics[trim={0 10 0 18},clip, width=0.49\linewidth ]{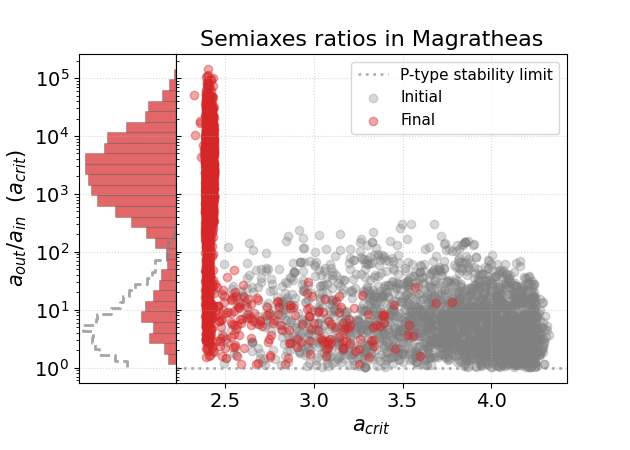}
        \includegraphics[trim={0 10 0 18},clip, width=0.49\linewidth ]{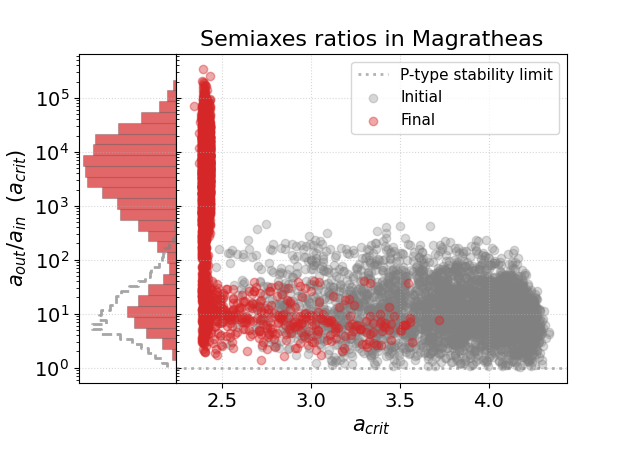} \\
        
        \includegraphics[trim={0 10 0 18},clip, width=0.49\linewidth ]{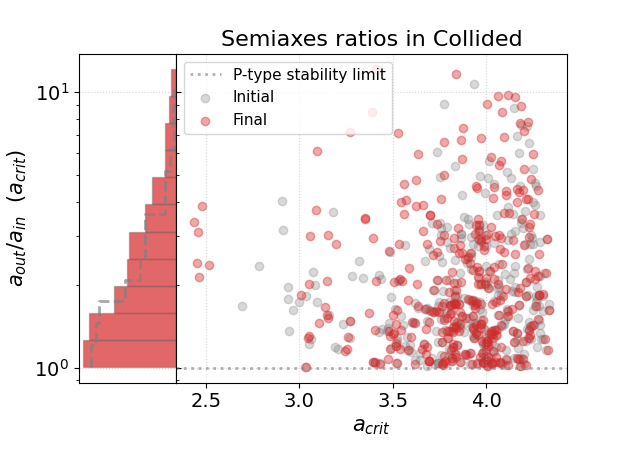} 
        \includegraphics[trim={0 10 0 18},clip, width=0.49\linewidth ]{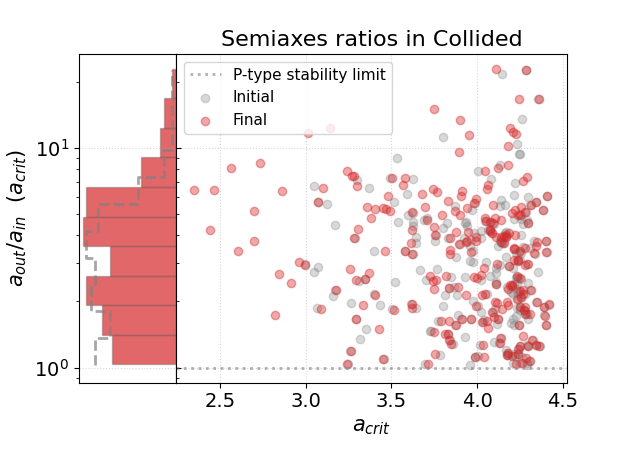} \\
    
        \includegraphics[trim={0 10 0 18},clip, width=0.48\linewidth]{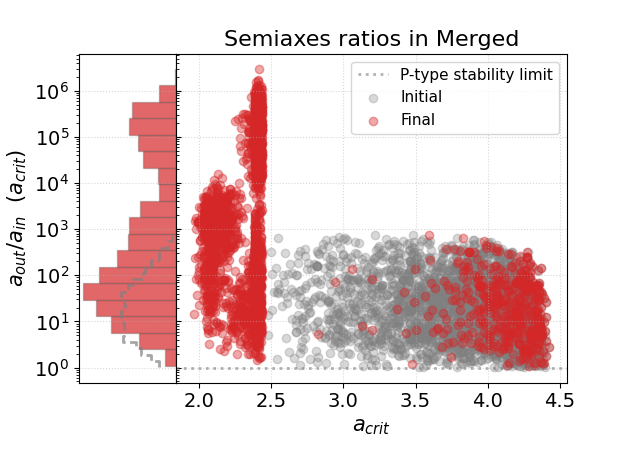}
        \includegraphics[trim={0 10 0 18},clip, width=0.48\linewidth]{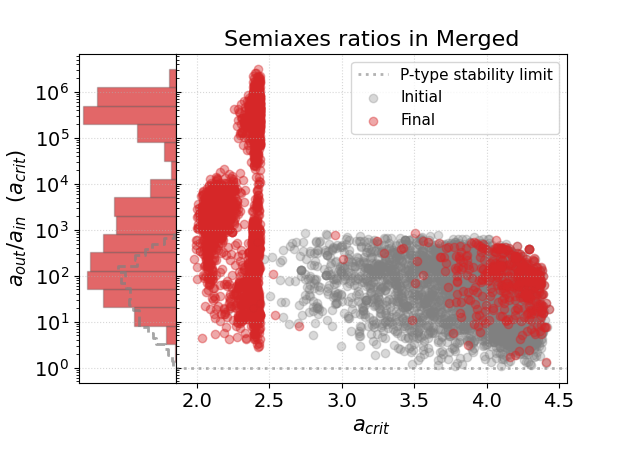} \\
    
        \includegraphics[trim={0 10 0 18},clip, width=0.48\linewidth]{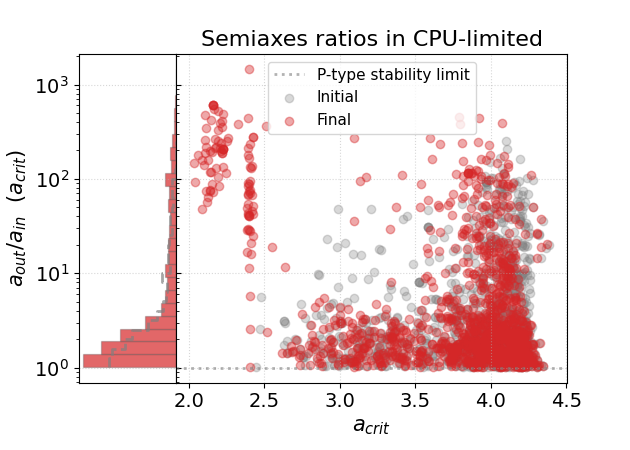} 
        \includegraphics[trim={0 10 0 18},clip, width=0.48\linewidth]{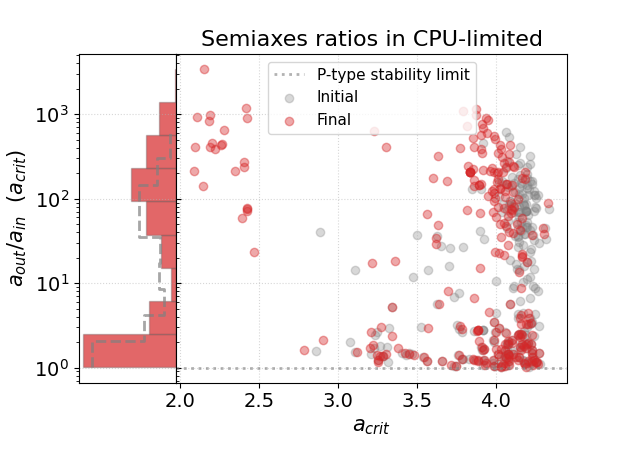}
        
        \caption{semi-major axes ratios of outer to inner binary as a function of the critical ratio, for each system. $ a_{crit} $ depends only on the inner binary, as shown in \secname~\ref{subsec:stability}. 
        On the left side of each plot, a histogram of the ratios distributions: please note that the logarithmic bins do alter the peaks shape, which in linear scale would always fall towards the lowest values.
        \emph{Left}: Pop. A, \emph{right}: Pop. B.}
        \label{fig:semiaxes_ratios}
    \end{figure*}

\end{appendix}


\end{document}